\pdfoutput=1
\documentclass[twocolumn,trackchanges]{aastex61}

\usepackage{graphicx}

\received{---}
\revised{---}
\accepted{---}

\shorttitle{Northern Eclipsing Algol-Type systems in CSS}
\shortauthors{Papageorgiou et al.}

\begin{document}

\title{An Updated Catalog of 4680 Northern Eclipsing Binaries with Algol-Type LIGHT CURVE MORPHOLOGY in the Catalina Sky Surveys}

\correspondingauthor{Athanasios Papageorgiou}
\email{apapageo@astro.puc.cl}

\author{Athanasios Papageorgiou} 
\affiliation{Pontificia Universidad Cat\'olica de Chile, Facultad de F\'{i}sica, Instituto de Astrof\'\i sica, Av. Vicu\~{n}a Mackenna 4860, 782-0436 Macul, Santiago, Chile}
\affiliation{Millennium Institute of Astrophysics, Santiago, Chile}

\author{M\'arcio Catelan} 
\altaffiliation{On sabbatical leave at The Observatories of the Carnegie Institution for Science, 813 Santa Barbara Street, Pasadena, CA 91101, USA}
\affiliation{Pontificia Universidad Cat\'olica de Chile, Facultad de F\'{i}sica, Instituto de Astrof\'\i sica, Av. Vicu\~{n}a Mackenna 4860, 782-0436 Macul, Santiago, Chile}
\affiliation{Millennium Institute of Astrophysics, Santiago, Chile}

\author{Panagiota-Eleftheria Christopoulou}
\affiliation{Department of Physics, University of Patras, 26500, Patra, Greece}

\author{Andrew J. Drake}
\affiliation{California Institute of Technology, 1200 East California,  Boulevard, CA 
91225, USA}

\author{S.G. Djorgovski}
\affiliation{California Institute of Technology, 1200 East California,  Boulevard, CA 
91225, USA}

\begin{abstract}
We present an updated catalog of 4680 northern eclipsing binaries (EBs) with Algol-type light curve morphology (i.e., with well-defined beginning and end of primary and secondary eclipses), using data from the Catalina Sky Surveys. Our work includes revised period determinations, phenomenological parameters of the light curves, and system morphology classification based on machine learning techniques. While most of the new periods are in excellent agreement with those provided in the original Catalina catalogs, improved values are now available for $\sim 10\%$ of the stars. A total of 3456 EBs were classified as detached and 449 as semi-detached, while 145 cannot be classified unambiguously into either subtype. 
The majority of the SD systems seems to be comprised of short-period Algols. 
By applying color criteria, we searched for K- and M-type dwarfs in these data, and present a subsample of 609 EB candidates for further investigation. We report 119 EBs (2.5$\%$ of the total sample) that show maximum quadrature light variations over long timescales, with periods bracketing the range $4.5-18$~yrs and fractional luminosity variance of $0.04-0.13$. We discuss possible causes for this,  making use of models of variable starspot activity in our interpretation of the results.


\end{abstract}


\section{Introduction} \label{sec:intro}

Our thinking about eclipsing binary stars (EBs) has undergone a tremendous change in the last decade. EBs are one of nature's best laboratories for determining the fundamental physical properties of stars, and thus for testing the predictions of theoretical models (e.g., Torres et al. 2010; Catelan \& Smith 2015, and references therein). A large number of eclipsing Algol-type (EA) binaries, for which the beginning and end of eclipses are well defined,  have been discovered recently as a by-product of several wide-field, ground-based photometric surveys, some of which are dedicated to the detection of variable stars. Among these surveys, one finds the Catalina Sky Survey (CSS, Larson et al. 2003), the Visible and Infrared Survey Telescope for Astronomy (VISTA) Variables in the Via Lactea (VVV, Minniti et al. 2010; Catelan et al. 2013), the asteroid survey LINEAR (Stokes et al. 2000; Palaversa et al. 2013), and the All Sky Automated Survey (ASAS, Pojmanski, 1997; Pojmanski et al. 2005), the Northern Sky Variability Survey (NSVS, Wo{\'z}niak et al. 2004), the Transatlantic Exoplanet Survey (TrES, Alonso et al. 2004; Alonso et al. 2007), the Optical Gravitational Lensing Experiment (OGLE, Udalski et al. 1992) survey, the Hungarian-made Automated Telescope Network exoplanet survey (HATNet, Bakos et al. 2004), and the Wide Angle Search for Planets (SuperWASP, Pollacco et al. 2006; Christian et al. 2006), among others \citep[see][for recent reviews and references]{gk17,is17}. 

On the basis of light curve (LC) morphology, EA-type eclipsing systems, with clearly defined eclipses on their LCs, include both (D) detached and semi-detached (SD) systems. As a rule, in order to establish the actual system configuration of any individual EB having such an Algol-type light curve morphology, a detailed physical modeling is required.

 The aforementioned projects are very useful to understand the photometric properties of the different types of binaries, affording for instance statistical studies of the properties of EA systems. In addition, large samples provide the opportunity for special cases of binaries that need dedicated follow-up observations to emerge, or even to reveal new classes (e.g., the Heartbeat stars, Welsh et al. 2011).

In this work, we use the northern data from CSS (which continues collecting data to this day) in order to complete our search for detached EBs and to present an updated and more detailed catalog of their properties, in comparison with \citet{adea09,adea14a,adea14b,adea17}. The additional data allow not only to revise their periods and class but also to derive the phenomenological and physical parameters of selected detached systems. Furthermore, we were able to search for systems exhibiting long-term variation, or which may harbor low-mass components. 

This paper, the first of a series on the subject, is organized as follows. In Section~\ref{sec:Observations}, we briefly describe the CSS data that we use in our analysis. The construction of the sample and an outline of the analysis methods are explained in Section~\ref{sec:EA}. Estimates of the periods and morphological features, and a physical classification of the EA type, are given in Section~\ref{subsec:Classification}. In Sections~\ref{subsec:Long term variations} and \ref{sec:Low mass EBs}, we discuss all the results, followed by a brief summary of our work in Section~\ref{sec:conclusion}.

\section{Observations} \label{sec:Observations}
Observations were carried out during 2004-2016 using the three telescopes of the Catalina Sky Surveys \cite{adea09},\footnote{\url{http://catalinadata.org}} covering the sky declination range $\delta = [-75, +65]$~deg, but avoiding crowded stellar regions within $10-15$~deg of the Galactic plane. The main goal of the survey is to discover Near-Earth Objects and Potential Hazardous Asteroids. Nevertheless, time-series photometry for $\sim 200$ million variable sources has been accumulated through CSS. In order to maximize the throughput, the observations are taken unfiltered, and the magnitudes  transformed to an approximate $V$ magnitude ($V_{\rm CSS}$; Drake et al. 2013). The photometry was performed using the aperture photometry program SExtractor (Bertin \& Arnouts 1996).

\begin{figure*}[htb!]
\plotone{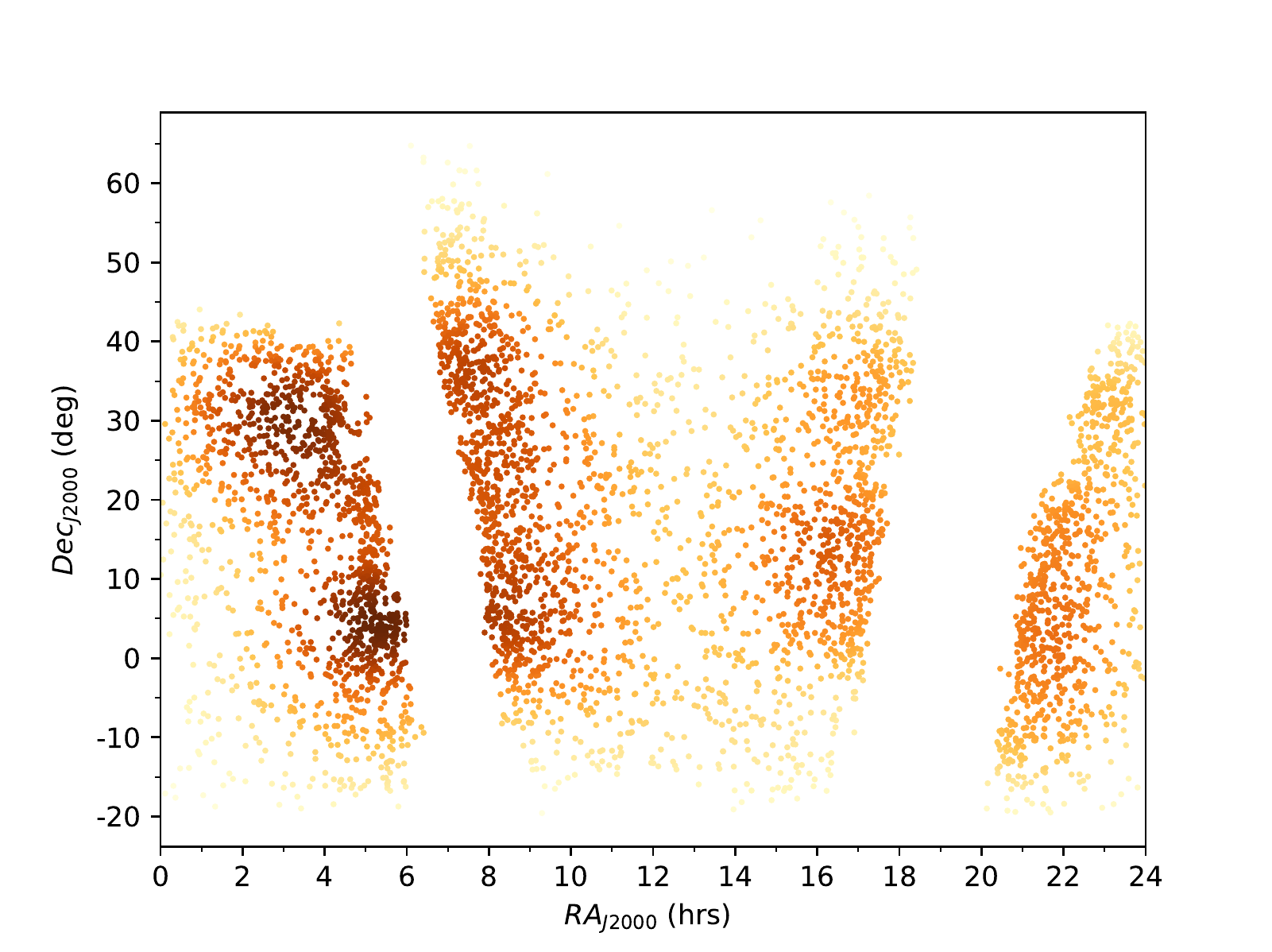}
\plotone{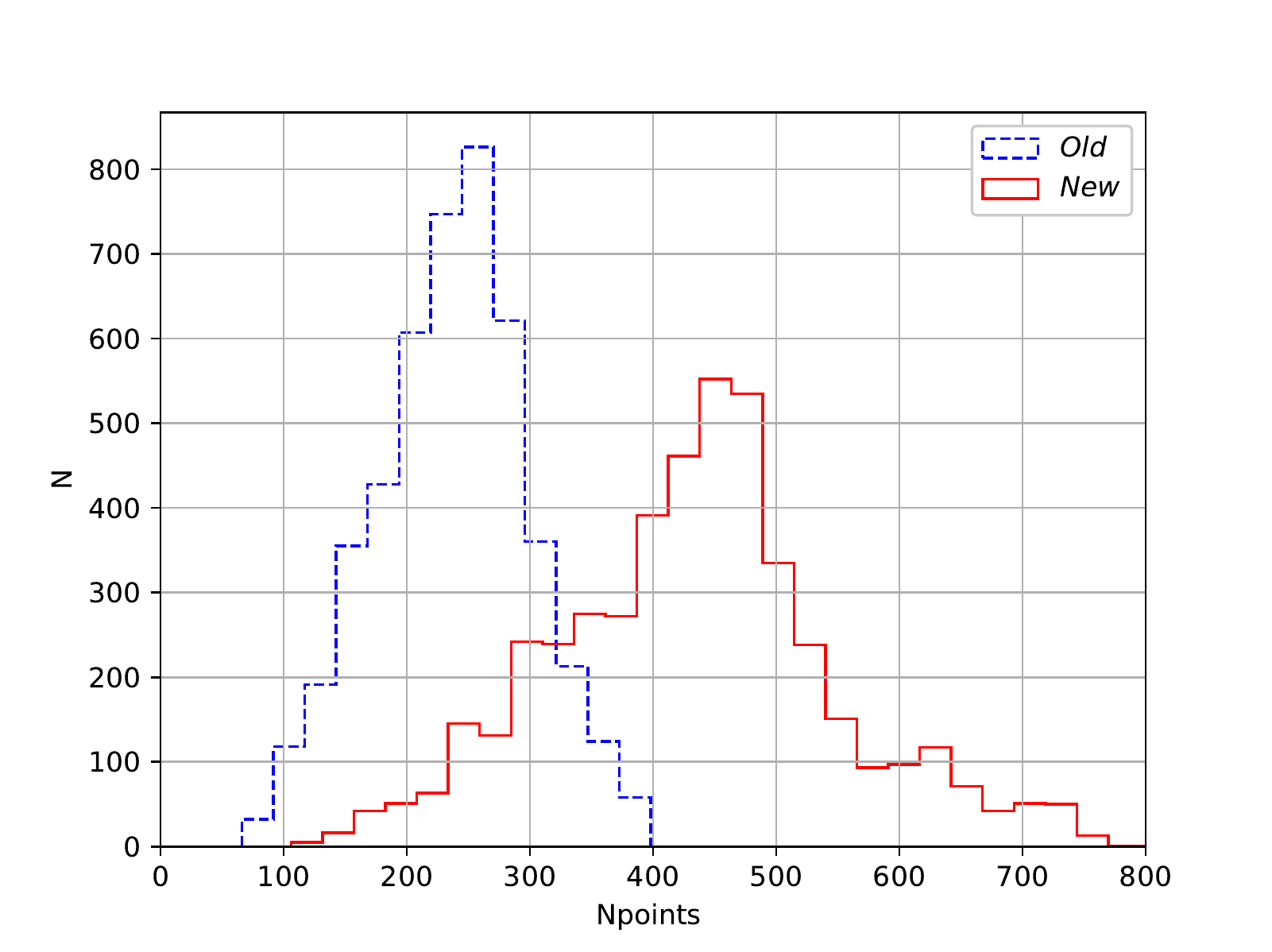}
\caption{{\em Top}: Sky distribution of 4683 EBs in the CSS catalog. {\em Bottom}: Distribution of the total number of photometric points per LC. The binary systems from the previous (Drake et al. 2014a) and new data releases (with the additional available data) are marked in the blue and red histogram, respectively. 
\label{fig:f1}}
\end{figure*}

In this study, we use Catalina Surveys Data Release 2\footnote{\url{http://nesssi.cacr.caltech.edu/DataRelease/}} (CSDR2) with additional, not publicly accessible data, spanning 12 years (2004-2016). We focus on the sample of Drake et al. (2014a) of 4683 eclipsing binaries originally classified as EA type on the basis of 8 years of data. These cover a region of right ascension (RA) between 0 and 24~hours and declination (Dec) between $-22$ and $+65$~deg, as shown in Figure~\ref{fig:f1} ($\it{top}$). The bottom panel of Figure~\ref{fig:f1} shows the distribution of the new available data used in the present paper against the CSDR2 data. The total number of photometric points for the  candidate systems that we studied significantly exceeds that available in the previous release.

\begin{figure*}[htb!]
\plotone{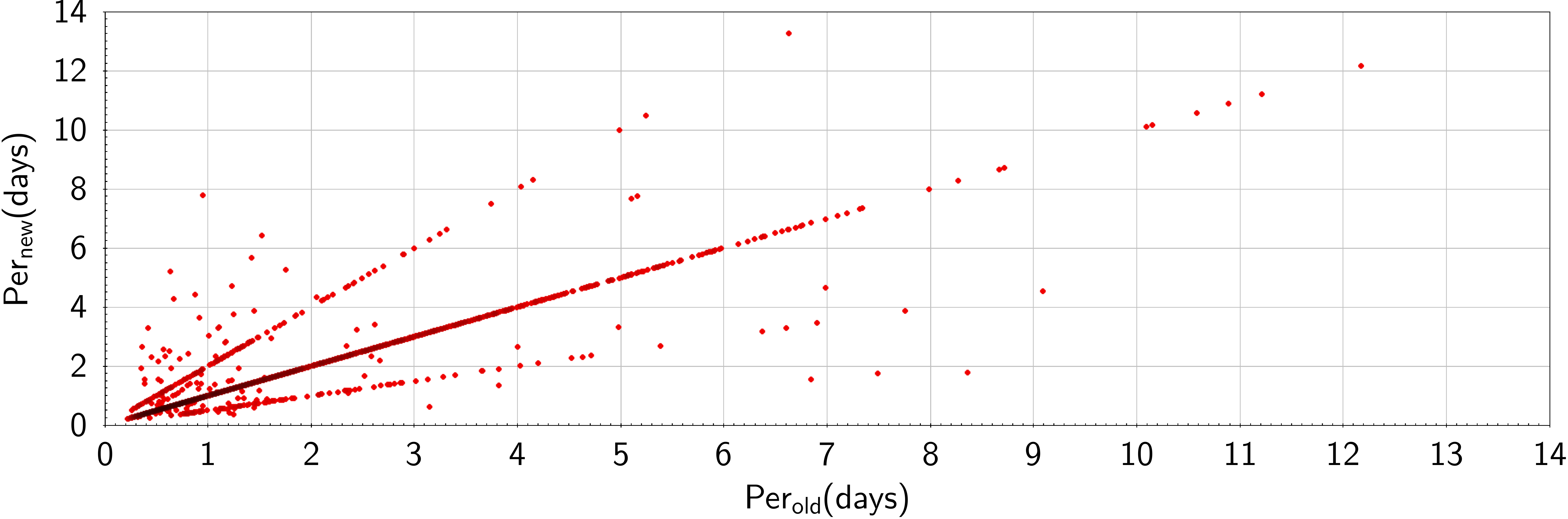}
\caption{Refined periods of 4680 EB stars. While most of the new periods ($Per_{\rm new})$ are in excellent agreement with those provided in Drake et al. (2014a, $Per_{\rm old}$), improved values are now available for $\sim 10\%$ of the stars. The differences are mostly due to aliases. \label{fig2:f13}}
\end{figure*}

\startlongtable
\begin{longrotatetable}
\begin{deluxetable*}{CCCCCCCCCCC}
\tablecaption{CRTS EA systems \label{Tab:T1a}}
\tablewidth{700pt}
\tabletypesize{\scriptsize}
\tablehead{
\colhead{Name} & \colhead{ID} & \colhead{RA} & \colhead{Dec} 
& \colhead{MJD\tablenotemark{a}} & \colhead{Per}
& \colhead{$<V_{err}>$\tablenotemark{b}}  & \colhead{Npoints}& \colhead{Class} & \colhead{LMCand} & \colhead{LongTerm}   \\ 
\colhead{} & \colhead{} & \colhead{(h:m:s)} & \colhead{($\arcdeg$ : $\arcmin$ : $\arcsec$)} 
& \colhead{(days)} & \colhead{(days)}
& \colhead{(mag)}  & \colhead{}& \colhead{} & \colhead{} 
}
\startdata
CSS$\_$J235945.5+303731  &  11291130655  &  23:59:45.5  &  +30:37:31.8  &  54265.53875  &  2.68651  &  0.0138  &  391  &  D  &  -  &  -  \\ 
CSS$\_$J235856.7+371823  &  11381030196  &  23:58:56.7  &  +37:18:23.5  &  55062.35518  &  1.35464  &  0.0136  &  312  &  D  &  -  &  -  \\ 
CSS$\_$J235816.7+293325  &  11291130352  &  23:58:16.7  &  +29:33:25.3  &  53537.41966  &  0.72949  &  0.0457  &  390  &  SD  &  -  &  -  \\ 
CSS$\_$J235756.9-023247  &  10011280051  &  23:57:56.9  &  -02:32:47.2  &  54747.25303  &  1.74457  &  0.0133  &  325  &  N/A  &  -  &  -  \\ 
CSS$\_$J235715.5+305455  &  11291130739  &  23:57:15.5  &  +30:54:55.4  &  53563.82135  &  2.84991  &  0.0157  &  381  &  D  &  -  &  -  \\ 
CSS$\_$J235538.3+384723  &  11381030654  &  23:55:38.3  &  +38:47:23.1  &  55508.59468  &  0.46792  &  0.0356  &  312  &  N/A  &  -  &  -  \\ 
CSS$\_$J235444.8+305751  &  11291130763  &  23:54:44.8  &  +30:57:51.9  &  54394.71148  &  0.82131  &  0.0279  &  349  &  D  &  -  &  -  \\ 
CSS$\_$J235401.4+374029  &  11381030304  &  23:54:01.4  &  +37:40:29.7  &  56558.31886  &  0.50473  &  0.0232  &  312  &  D  &  -  &  -  \\ 
CSS$\_$J235313.6-021850  &  10011280091  &  23:53:13.6  &  -02:18:50.2  &  53655.26418  &  0.50952  &  0.0152  &  325  &  D  &  -  &  -  \\ 
CSS$\_$J235227.0+395515  &  11400990133  &  23:52:27.0  &  +39:55:15.3  &  53694.94006  &  1.5311  &  0.0287  &  218  &  D  &  -  &  -  \\ 
CSS$\_$J235151.3+035409  &  11041280214  &  23:51:51.3  &  +03:54:09.0  &  55850.15160  &  2.98858  &  0.0136  &  375  &  D  &  -  &  -  \\ 
CSS$\_$J235104.0+115651  &  11121260136  &  23:51:04.0  &  +11:56:51.3  &  54095.14156  &  0.81240  &  0.0138  &  442  &  D  &  -  &  -  \\ 
CSS$\_$J234952.4-012059  &  10011280269  &  23:49:52.4  &  -01:20:59.4  &  53637.21180  &  1.76549  &  0.0390  &  325  &  D  &  -  &  -  \\ 
CSS$\_$J234939.6-004257  &  10011280381  &  23:49:39.6  &  -00:42:57.1  &  55113.21426  &  0.49138  &  0.0345  &  324  &  D  &  -  &  -  \\ 
CSS$\_$J234850.3+133300  &  11121260465  &  23:48:50.3  &  +13:33:00.1  &  55858.22093  &  1.46550  &  0.0288  &  442  &  D  &  -  &  -  \\ 
CSS$\_$J234828.2+403240  &  11400990337  &  23:48:28.2  &  +40:32:40.1  &  55119.12598  &  0.93421  &  0.0182  &  219  &  N/A  &  -  &  -  \\ 
CSS$\_$J234827.2+392032  &  11381030834  &  23:48:27.2  &  +39:20:32.8  &  55348.41153  &  1.70451  &  0.0172  &  308  &  N/A  &  -  &  -  \\ 
CSS$\_$J234826.5+271203  &  11261160472  &  23:48:26.5  &  +27:12:03.6  &  54732.19943  &  0.86780  &  0.0237  &  371  &  D  &  -  &  -  \\ 
CSS$\_$J234819.9+344833  &  11351060279  &  23:48:19.9  &  +34:48:33.9  &  55943.75582  &  2.52099  &  0.0130  &  326  &  N/A  &  -  &  -  \\ 
CSS$\_$J234734.4+203331  &  11211200175  &  23:47:34.4  &  +20:33:31.9  &  55088.41305  &  0.86457  &  0.0144  &  429  &  D  &  -  &  -  \\ 
CSS$\_$J234700.0+180015  &  11181220236  &  23:47:00.0  &  +18:00:15.6  &  54480.10645  &  3.07628  &  0.0169  &  431  &  D  &  -  &  -  \\ 
CSS$\_$J234554.3-003131  &  10011270462  &  23:45:54.3  &  -00:31:31.1  &  54477.07608  &  0.69255  &  0.0234  &  398  &  D  &  -  &  -  \\ 
CSS$\_$J234502.5+415419  &  11400980842  &  23:45:02.5  &  +41:54:19.9  &  54632.41237  &  0.85779  &  0.0763  &  245  &  D  &  -  &  -  \\ 
CSS$\_$J234439.7+055255  &  11071260052  &  23:44:39.7  &  +05:52:55.8  &  56301.08055  &  0.50740  &  0.0136  &  407  &  SD  &  -  &  -  \\ 
CSS$\_$J234348.2+270630  &  11261150467  &  23:43:48.2  &  +27:06:30.4  &  55024.49074  &  2.15783  &  0.0135  &  438  &  D  &  -  &  -  \\ 
CSS$\_$J234339.1+362901  &  11351050807  &  23:43:39.1  &  +36:29:01.2  &  55366.42500  &  0.31764  &  0.1027  &  386  &  D  &  -  &  -  \\ 
CSS$\_$J234331.2-010354  &  10011270364  &  23:43:31.2  &  -01:03:54.4  &  54009.17422  &  1.13869  &  0.0147  &  398  &  D  &  -  &  -  \\ 
CSS$\_$J234306.1+060347  &  11071260081  &  23:43:06.1  &  +06:03:47.7  &  54730.34298  &  3.6267  &  0.0335  &  407  &  D  &  -  &  -  \\ 
CSS$\_$J234230.6+410139  &  11400980526  &  23:42:30.6  &  +41:01:39.8  &  54394.14236  &  0.52837  &  0.0256  &  248  &  D  &  -  &  -  \\ 
CSS$\_$J234137.0+264453  &  11261150363  &  23:41:37.0  &  +26:44:53.2  &  55009.56074  &  0.33624  &  0.1115  &  436  &  D  &  -  &  -  \\ 
CSS$\_$J234116.3+392234  &  11381020853  &  23:41:16.3  &  +39:22:34.6  &  53655.20953  &  0.70380  &  0.0213  &  325  &  D  &  -  &  -  \\ 
\enddata
\tablenotetext{a}{Epoch at primary minimum }
\tablenotetext{b}{Mean photometric error ($V_{CSS}$) }
\end{deluxetable*}
\end{longrotatetable}
 
\startlongtable
\begin{longrotatetable}
\begin{deluxetable*}{ccccccccccc}
\tablecaption{Phenomenological parameters of 4680 EBs \label{Tab:T1b}}
\tablehead{
\colhead{Name} & \colhead{Amp} & \colhead{$Amp_{err}$\tablenotemark{a}} & \colhead{MinI} & \colhead{$MinI_{err}$\tablenotemark{a}} & \colhead{MinII}& \colhead{$MinII_{err}$\tablenotemark{a}} & \colhead{MaxI}& \colhead{$MaxI_{err}$\tablenotemark{a}} & \colhead{$|MinI-MinII|$} & \colhead{$|MinI-MinII|_{err}$\tablenotemark{a}}  \\ 
\colhead{} & \colhead{(mag)} & \colhead{(mag)}  & \colhead{(mag)} & \colhead{(mag)} & \colhead{(mag)} & \colhead{(mag)} & \colhead{(mag)} & \colhead{(mag)} & \colhead{(mag)} & \colhead{(mag)} 
}
\startdata
CSS$\_$J235945.5+303731  &  1.9357  &  0.0225  &  15.6800  &  0.0205  &  13.8708  &  0.0130  &  13.7176  &  0.0145  &  1.8092  &  0.0216  \\ 
CSS$\_$J235856.7+371823  &  0.2647  &  0.0083  &  13.5267  &  0.0061  &  13.3597  &  0.0069  &  13.2358  &  0.0062  &  0.1669  &  0.0088  \\ 
CSS$\_$J235816.7+293325  &  1.4841  &  0.0276  &  18.2052  &  0.0231  &  16.9536  &  0.0242  &  16.6789  &  0.0180  &  1.2516  &  0.0321  \\ 
CSS$\_$J235756.9-023247  &  0.6316  &  0.0344  &  13.3882  &  0.0269  &  13.2883  &  0.0288  &  12.7309  &  0.0239  &  0.0999  &  0.0379  \\ 
CSS$\_$J235715.5+305455  &  0.3071  &  0.0087  &  14.6847  &  0.0082  &  14.6709  &  0.0076  &  14.3497  &  0.0031  &  0.0137  &  0.0111  \\ 
CSS$\_$J235538.3+384723  &  0.2798  &  0.0150  &  16.6196  &  0.0117  &  16.4139  &  0.0152  &  16.3397  &  0.0104  &  0.2057  &  0.0192  \\ 
CSS$\_$J235444.8+305751  &  0.8093  &  0.0271  &  16.7195  &  0.0188  &  16.5878  &  0.0268  &  15.8768  &  0.0214  &  0.1317  &  0.0316  \\ 
CSS$\_$J235401.4+374029  &  0.3562  &  0.0283  &  15.7998  &  0.0205  &  15.7252  &  0.0240  &  15.4119  &  0.0217  &  0.0745  &  0.0301  \\ 
CSS$\_$J235313.6-021850  &  0.7300  &  0.0209  &  15.0009  &  0.0173  &  14.8783  &  0.0214  &  14.2432  &  0.0132  &  0.1225  &  0.0268  \\ 
CSS$\_$J235227.0+395515  &  0.7084  &  0.0145  &  16.6152  &  0.0112  &  16.0845  &  0.0100  &  15.8725  &  0.0103  &  0.5306  &  0.0143  \\ 
CSS$\_$J235151.3+035409  &  1.0865  &  0.0186  &  14.3761  &  0.0159  &  13.3307  &  0.0121  &  13.2895  &  0.0107  &  1.0454  &  0.0200  \\ 
CSS$\_$J235104.0+115651  &  0.3206  &  0.0128  &  14.0039  &  0.0105  &  13.9033  &  0.0130  &  13.6566  &  0.0079  &  0.1006  &  0.0164  \\ 
CSS$\_$J234952.4-012059  &  1.0510  &  0.0264  &  17.5661  &  0.0219  &  16.7182  &  0.0194  &  16.4766  &  0.0165  &  0.8479  &  0.0283  \\ 
CSS$\_$J234939.6-004257  &  0.4143  &  0.0217  &  16.7682  &  0.0166  &  16.4422  &  0.0153  &  16.3179  &  0.0161  &  0.3259  &  0.0211  \\ 
CSS$\_$J234850.3+133300  &  0.5252  &  0.0190  &  16.4847  &  0.0175  &  16.2129  &  0.0172  &  15.9255  &  0.0081  &  0.2718  &  0.0243  \\ 
CSS$\_$J234828.2+403240  &  0.2586  &  0.0110  &  15.1143  &  0.0091  &  15.1082  &  0.0153  &  14.8268  &  0.0069  &  0.0061  &  0.0175  \\ 
CSS$\_$J234827.2+392032  &  0.3341  &  0.0140  &  14.9943  &  0.0120  &  14.7398  &  0.0562  &  14.6316  &  0.0079  &  0.2544  &  0.0574  \\ 
CSS$\_$J234826.5+271203  &  0.4481  &  0.0109  &  15.9527  &  0.0088  &  15.8867  &  0.0097  &  15.4728  &  0.0072  &  0.0659  &  0.0128  \\ 
CSS$\_$J234819.9+344833  &  1.3043  &  0.0420  &  12.7552  &  0.0366  &  11.8263  &  0.0086  &  11.4509  &  0.0230  &  0.9289  &  0.0376  \\ 
CSS$\_$J234734.4+203331  &  0.7597  &  0.0145  &  14.8161  &  0.0123  &  14.3558  &  0.0187  &  14.0290  &  0.0083  &  0.4602  &  0.0222  \\ 
CSS$\_$J234700.0+180015  &  0.3179  &  0.0088  &  14.9473  &  0.0072  &  14.9158  &  0.0085  &  14.6010  &  0.0056  &  0.0315  &  0.0109  \\ 
CSS$\_$J234554.3-003131  &  0.8222  &  0.0161  &  16.3255  &  0.0132  &  15.8831  &  0.0151  &  15.4719  &  0.0103  &  0.4424  &  0.0195  \\ 
CSS$\_$J234502.5+415419  &  0.6371  &  0.0464  &  18.1580  &  0.0394  &  18.0035  &  0.0285  &  17.4683  &  0.0268  &  0.1544  &  0.0473  \\ 
CSS$\_$J234439.7+055255  &  0.4433  &  0.0165  &  13.5755  &  0.0130  &  13.3442  &  0.0142  &  13.1060  &  0.0112  &  0.2313  &  0.0187  \\ 
CSS$\_$J234348.2+270630  &  0.5531  &  0.0119  &  13.7762  &  0.0108  &  13.7734  &  0.0200  &  13.1970  &  0.0054  &  0.0028  &  0.0226  \\ 
CSS$\_$J234339.1+362901  &  0.7290  &  0.0426  &  18.6783  &  0.0311  &  18.3661  &  0.0364  &  17.8895  &  0.0323  &  0.3121  &  0.0458  \\ 
CSS$\_$J234331.2-010354  &  0.1887  &  0.0051  &  14.3634  &  0.0045  &  14.2967  &  0.0074  &  14.1473  &  0.0025  &  0.0667  &  0.0086  \\ 
CSS$\_$J234306.1+060347  &  0.8140  &  0.0196  &  17.0581  &  0.0188  &  16.6652  &  0.0158  &  16.2079  &  0.0060  &  0.3928  &  0.0244  \\ 
CSS$\_$J234230.6+410139  &  0.3710  &  0.0140  &  16.0309  &  0.0107  &  15.7939  &  0.0117  &  15.6270  &  0.0100  &  0.2369  &  0.0153  \\ 
CSS$\_$J234137.0+264453  &  0.7832  &  0.0390  &  18.7695  &  0.0323  &  18.4569  &  0.0328  &  17.9221  &  0.0261  &  0.3125  &  0.0438  \\ 
CSS$\_$J234116.3+392234  &  0.6616  &  0.0150  &  15.8778  &  0.0119  &  15.3690  &  0.0106  &  15.1853  &  0.0101  &  0.5088  &  0.0154  \\ 
\enddata
\tablenotetext{a}{Estimated from the fitting}
\end{deluxetable*}
\end{longrotatetable}

\section{Identification of Algol-type Eclipsing Binaries in Catalina Sky Survey}  \label{sec:EA}
As we wanted to take all good data points of a light curve into account to search for periodic signals, we first cleaned the 4683 LCs of the initial sample. For every light curve a sigma clipping cut-off algorithm was used to discard erroneous data points with values outside the interval of $\pm 5 \sigma$ of the median relative flux, where $\sigma$ denotes the standard deviation computed from the whole light curve. Furthermore, by adopting a pre-define period from Drake et al. (2014a) we performed $5 \sigma$ clipping from the median value of each phase bin. Therefore, we avoid rejecting data points corresponding to an eclipse and we ensure that the data points with error larger than $5 \sigma$ are discarded as outliers, presumably due to unreliable measurements.

\subsection{Period search} \label{subsec:Periods}
\begin{figure*}[htb!]
\plotone{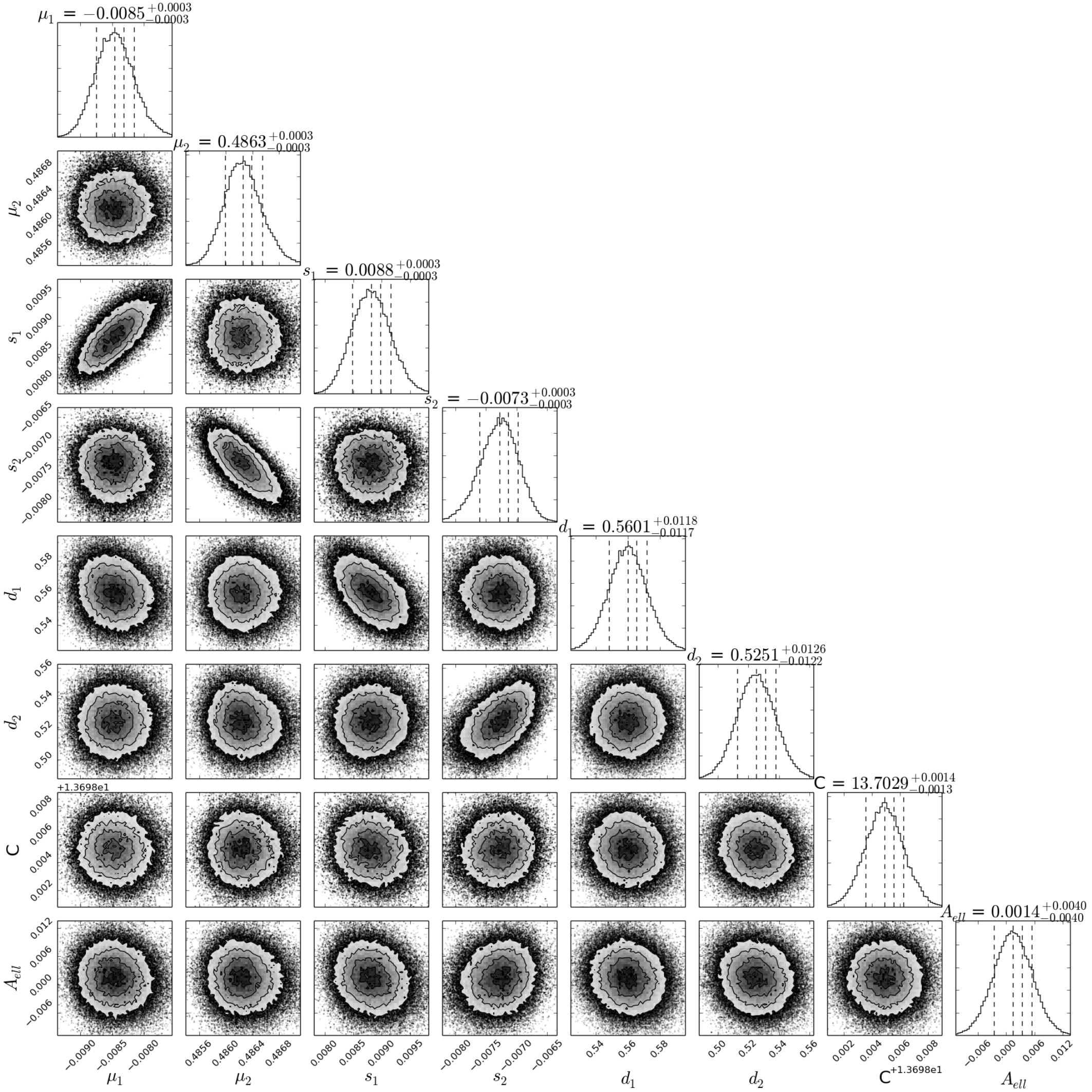}
\plottwo{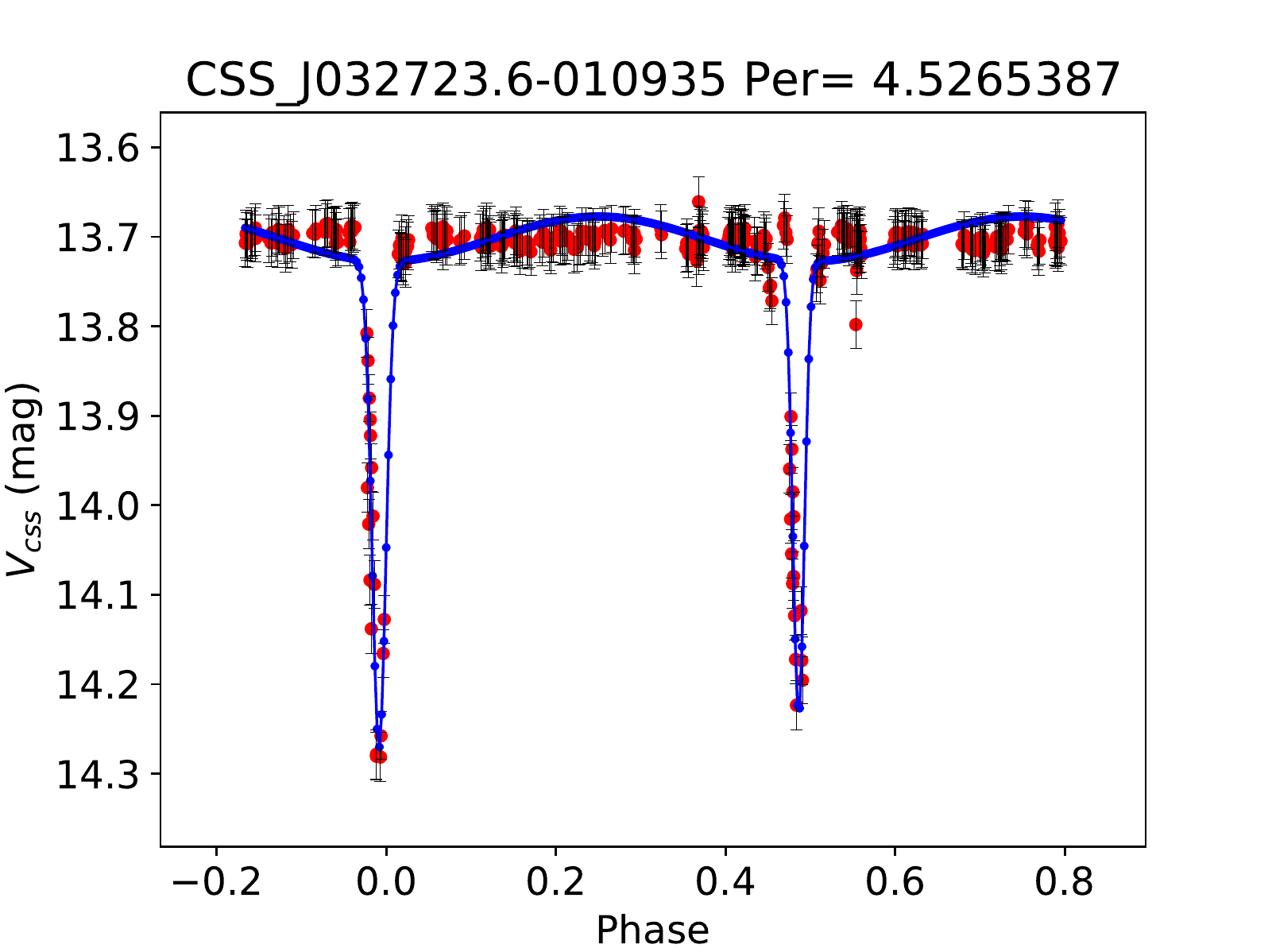}{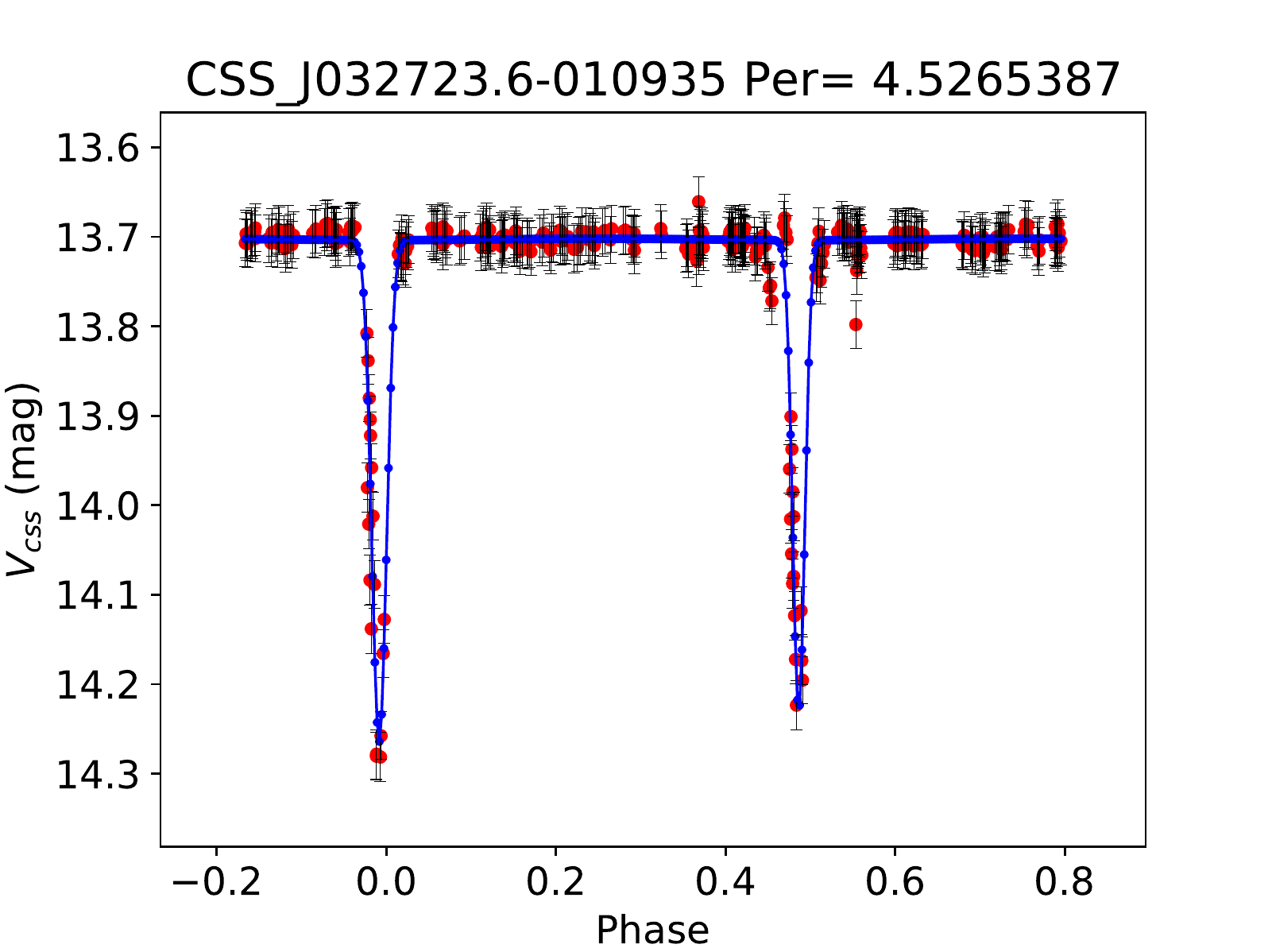}
\caption{{\em Top}:  One and two dimensional projections of the posterior probability distributions \citep{2014zndo.soft11020F} of a few parameters inferred from the TGM on each light curve. {\em Bottom}: Example light curve with the initial ({\em left}) and the final ({\em right}) TGM fitting coupled by MCMC. Blue dots and solid lines refer to  the resulting TGM, while red dots refer to  the CSS data. CSS IDs and periods are given on top of each light curve. \label{fig:f2}}
\end{figure*}

After cleaning and checking  a certain number of the resultant LCs, we applied a series  of period-finding  methods,  such as:

\begin{itemize}
\item Analysis of Variance (AoV, Schwarzenberg-Czerny 1989, and Devor 2005);
\item Box-Least Squares (BLS, Kov\'acs et  al. 2002);
\item Generalized Lomb-Scargle (GLS, Zechmeister and K\"urster 2009, Press et al. 1992);
\item Phase Dispersion Minimization (PDM, Stellingwerf 1978);
\item Correntropy  Kernelized  Periodogram (CKP, Protopapas et al. 2015).
\end{itemize}

\noindent The AoV, BLS, and GLS algorithms were applied through the command line utility VARTOOLS (Hartman \& Bakos 2016).

At first, the AoV method was applied, using a period range $[0.1-700]$~days and frequency resolution $0.1/T$ (where $T$ is the time span of the light curve), returning the top 5 peaks of the spectrum. The phase-folded LCs were visual inspected using these periods, and the best values were adopted. When the period values from AoV failed to phase-fold the light curves adequately, the other methods were applied, and the phased light curves were again visually inspected. A common issue encountered using periodograms, in the case of EBs, is the double/half period detection. For this reason, the majority of the phase-folded LCs were also examined using twice/half the detected period. 

Using the methodology described, we improved the period determination of the detached EB sample. While most of the new periods are in excellent agreement with those provided in Drake et al. (2014a), improved values are now available for $\sim 10\%$ of the 4680 stars  (Figure~\ref{fig2:f13}). Table~\ref{Tab:T1a} summarizes the results obtained for the latter, including also the mean photometric error ($V_{\rm err}$)\footnote{The original CSS photometric errors are significantly overestimated, as discussed in \citet{mgea17}. \citeauthor{mgea17} provide a corrective factor $f_{\rm corr}$ to compensate for this problem. The following analytical fit provides an excellent description of the data shown in Figure~1 of their paper:
\begin{equation}
f_{\rm corr} = a \, \left[1+\frac{(V-b)^2}{c^2 d}\right]^{-\left(\frac{d}{2}+\frac{1}{2}\right)},
\end{equation} 
\noindent with $a =  1.350$, $b = 19.491$, $c = 3.006$, and $d = 0.275$. The fit is valid between $V=14.0$ and 19.5~mag. For $V < 14.0$~mag, a value $f_{\rm corr} = 0.26$ is assumed; for $V >19.50$~mag, we adopt instead $f_{\rm corr} = 1.35$. All error values reported in this paper, including tables and plots, have been corrected according to this recipe.} and source coordinates (${\rm RA}_{\rm J2000}$, ${\rm Dec}_{\rm J2000}$). 

\begin{figure*}[htb!]
\plottwo{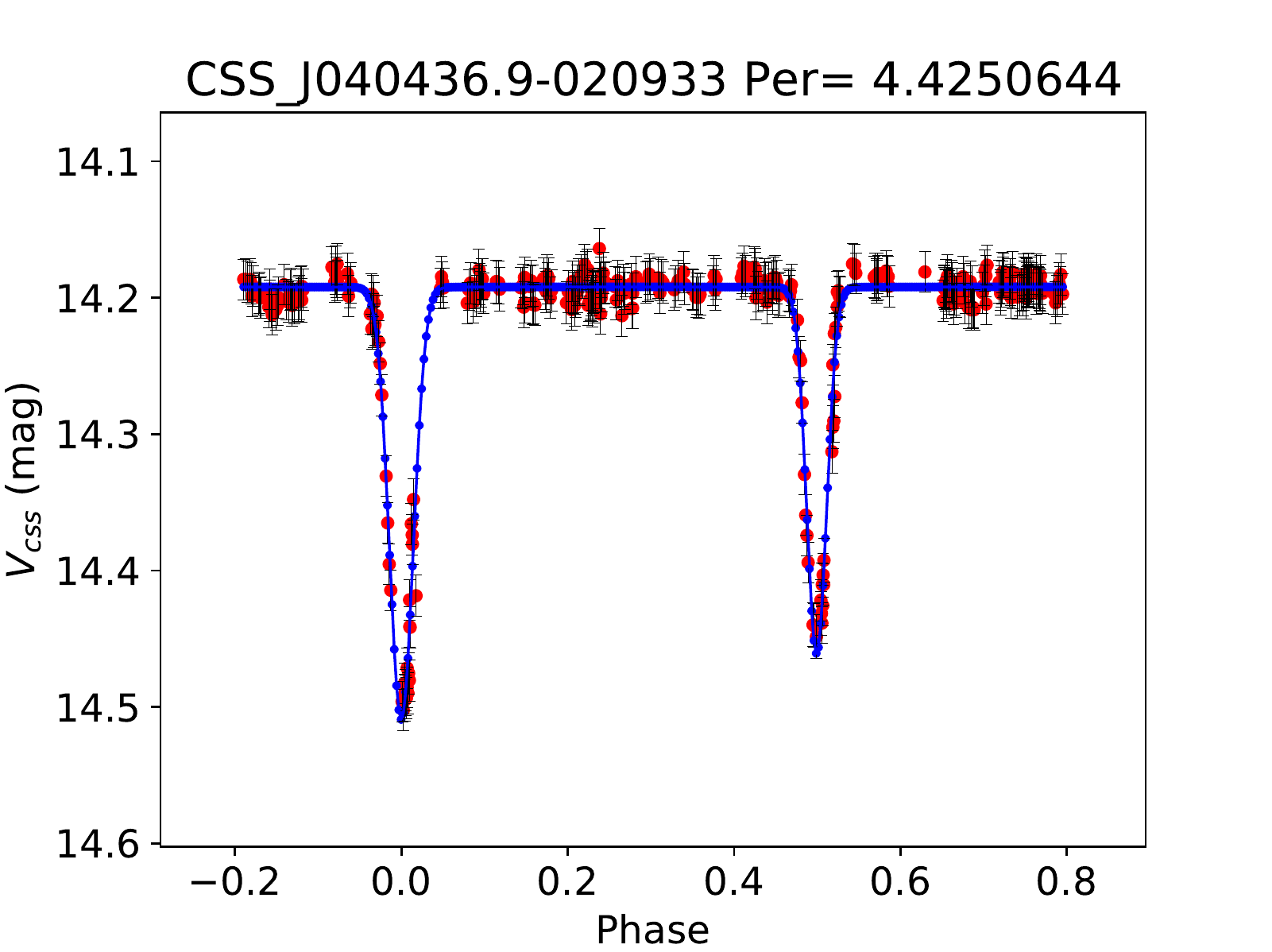}{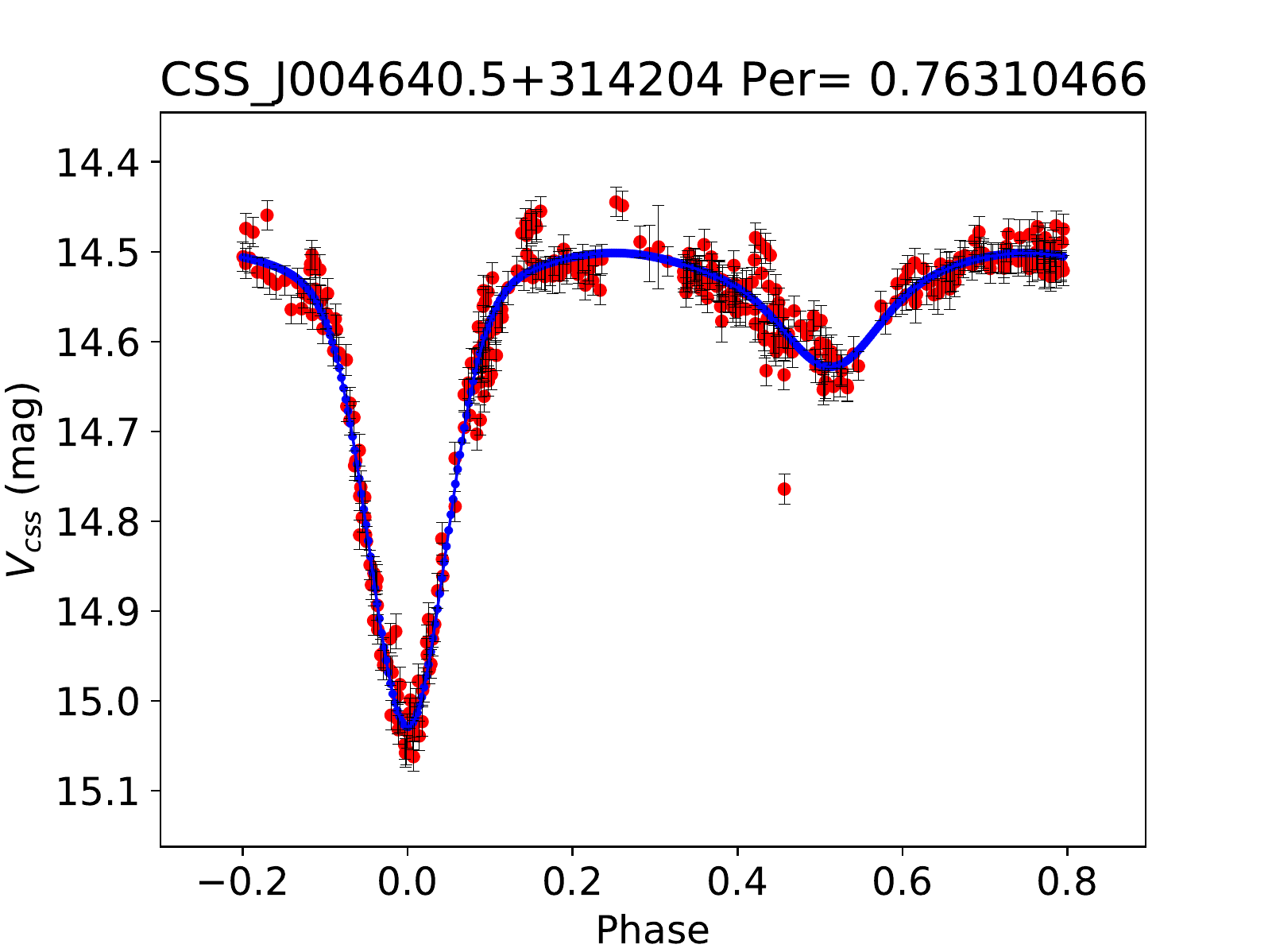}
\caption{Representative examples of a D system ({\em left}) and an SD system ({\em right}), obtained using TGM fitting. Symbols and colors as in Figure~\ref{fig:f2}. CSS IDs and periods are given on top of each light curve. \label{fig:f3}}
\end{figure*}

\subsection{Light curve phenomenological parameters} \label{subsec:phenomenological parameters}
After phase-folding the LCs as explained in the previous subsection, long-term variations were also removed, when present (see Section~\ref{subsec:Long term variations}). The light curve was then fitted using the LMFIT\footnote{\url{http://dx.doi.org/10.5281/zenodo.11813}} (Newville et al. 2016) module in Python\footnote{\url{http://www.python.org}} in order to derive its morphological features. Three different models were used, namely: a chain of second-order polynomials (Pr{\v s}a et al. 2008, Papageorgiou et al. 2014), Fourier series fitting and a Two-Gaussian Model (TGM, Mowlavi et al. 2017). The actual fitting process was overseen by the Levenberg-Marquardt (LM, Levenberg 1944; Marquardt 1963) nonlinear minimization algorithm.
We found that the TGM technique was much more robust and efficient, when applied to the stars in our sample. The procedure is based on modeling the geometry of LCs using Gaussian functions (to model the eclipses) and a cosine function (to model ellipsoidal variability, if present). Fitting a TGM to a time series is very sensitive to the adopted initial values of the parameters. We therefore used the LM parameter values as starting points on each Two-Gaussian Model. These include the phases ($\mu_{i}$), the half widths ($s_{i}$), and the depths ($d_{i}$) of the primary and secondary eclipses ($i=1,2$), the peak-to-peak amplitude of the ellipsoidal-like variation ($A_{\rm ell}$), and a constant ($C$) that equals the maximum light of the LC in the case of detached systems (Mowlavi et al. 2017, see their Figure~1).

\begin{figure*}[htb!]
\plotone{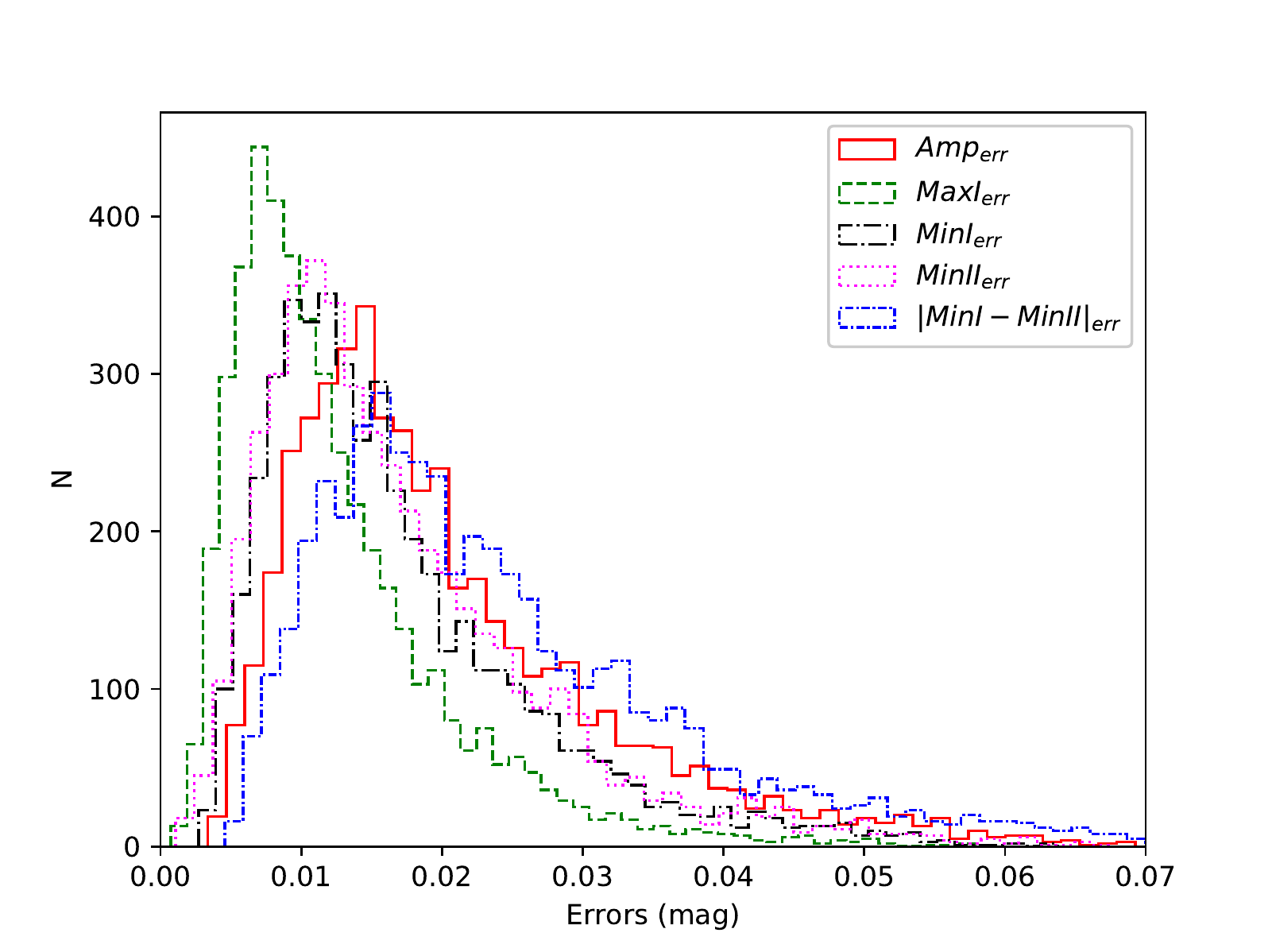}
\caption{Distribution of the obtained uncertainties in the parameters Amp, MaxI, MinI, MinII, and the difference $|{\rm MinI}-{\rm MinII}|$.
\label{fig:f8}}
\end{figure*}

Since the success of modeling the folded LCs depends on the time sampling, measurement uncertainties, initial guessing of eclipse locations, and additional intrinsic variability in one or both stars of the binary system, a Markov Chain Monte Carlo (MCMC) analysis was performed on each TGM of our LC sample, using the pyMC \citep{2015ascl.soft06005F} module\footnote{\url{ https://pymc-devs.github.io/pymc/}} in Python. The MCMC process begins by generating initial guesses for all the parameters randomly selected from a normal distribution  based on the final LM fitting parameter values and errors. The new fit is accepted or rejected using the Metropolis-Hastings algorithm \citep{1970Bimka..57...97H}, compared to the fitting carried out in the previous step. In order to avoid the biases that might be present in the initial solutions, the first 15,000 steps (of 200,000 steps in total) were discarded in the process. We then sampled this new synthetic model and discovered that the initial TGM model was noticeably different in some cases (Figure~\ref{fig:f2}). Examples of the folded LCs, classified as D and SD as discussed below, are presented in Figure~\ref{fig:f3}. The derived phenomenological parameters for 4680 EBs are presented in Table~\ref{Tab:T1b}. Such parameters include: the magnitude at primary (MinI) and secondary eclipse (MinII), the magnitude at maximum light out of the eclipses (MaxI), the difference between the eclipse depths (${\rm MinI}-{\rm MinII}$),  the amplitude (Amp) and the mean magnitude ($\langle V_{\rm mag}\rangle$). The histograms of the distribution of errors for the phenomenological parameters are shown in Figure~\ref{fig:f8}.  


\section{Classification} \label{subsec:Classification}
The EBs in previous CSS data releases were classified into D or SD systems based on visual inspection of the LCs. Lee (2015), using the Method for Eclipsing Component Identification (Devor et al. 2006), found 272 SD EBs amongst 2170 fitted LCs (of the total 4683), based on Roche lobe filling criteria.

Based on the system morphology classification, we performed, for the first time, an automated classification of the majority of EA-type CSS EBs with machine learning algorithms. In our search, unsupervised machine learning followed by a supervised learning was performed using 8000 synthetic LCs of D, SD, overcontact (OC), and ellipsoidal (ELL) EBs. As a training set, 2000 LCs were randomly selected for each class, out of a total of $\sim 32,\!000$ synthetic LCs. The synthetic LCs were created by a Monte Carlo-based script (Pr\v{s}a et al. 2008) in PHOEBE-scripter (Pr\v{s}a \& Zwitter 2005), using randomly selected parameters for each physical model. For each LC, 201 equally phased bins in the range [0,1] were utilized. 
 
\begin{figure*}[htb!]

\plotone{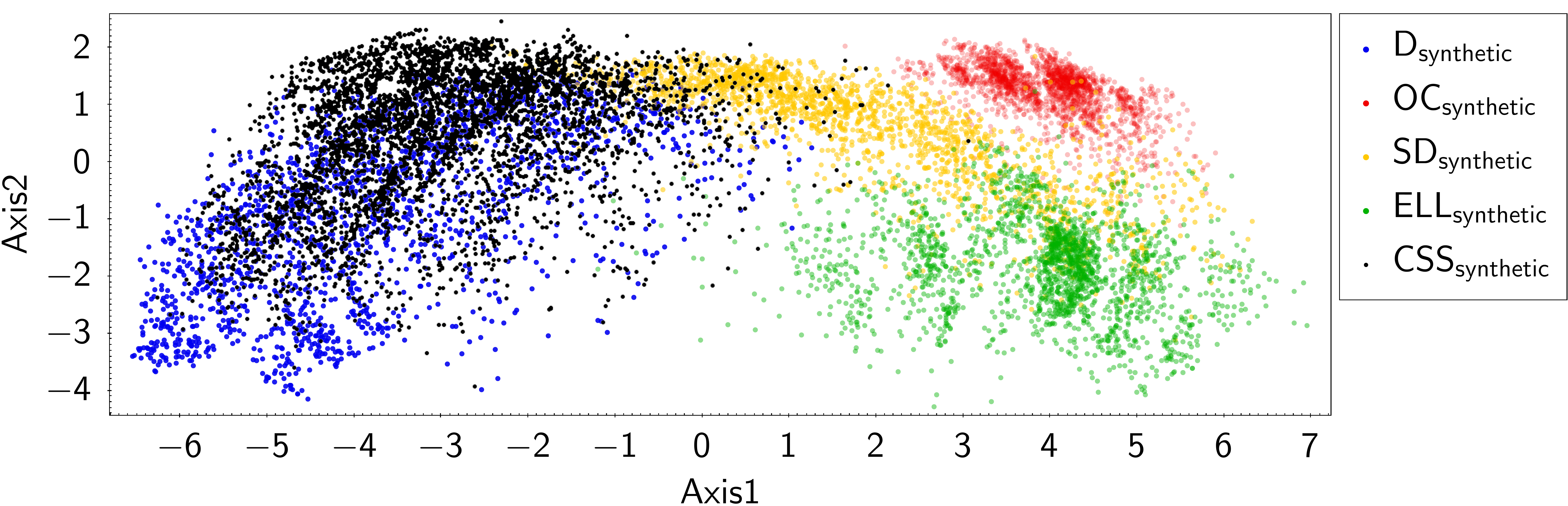}
\caption{Lower-dimensional input data space projection (2-D projection) applying the method of Isomap. The axes of the 2-D projection represent the top two eigenvectors of the geodesic distance matrix. Colored symbols indicate the distribution of synthetic LCs, whereas black dots indicate the positions of CSS sources. \label{fig:f4}}
\end{figure*}

An unsupervised learning was then performed for each of 4050 phenomenological models obtained from Section~\ref{subsec:phenomenological parameters}, selected according to fitting performance. This was done by applying a variety of methods through the scikit-learn\footnote{\url{http://scikit-learn.org/stable/index.html}} module (Pedregosa et al. 2012) in Python. Lower-dimensional space projections (2-D and 3-D) were found applying the method of complete isometric feature mapping with 160 nearest neighbours (Isomap, Tenenbaum et al. 2000) that separates the classes (Figure~\ref{fig:f4}). Considering the separation of the classes, we applied a supervised machine learning using the values of the 3-D projection as input for the training set and the CSS data. 
 
Furthermore, random Gaussian noise with a $\sigma =0.08$~mag was added to the sample of the training set and the phenomenological models. A variety of classifiers were applied, and we found that the best performance (validation score 92\%) was achieved by Support Vector Machine (SVM). However, similar validation scores ($\sim 89-91\%$) were achieved using Random Forest, Artificial Neural Network (ANN), and K-nearest neighbor (KN) classifiers.  

The confusion matrix of 2000 test synthetic EBs from the SVM classifier is presented in Table~\ref{Tab:confMat}. We found a discrepancy among the classifiers for 263 EBs. Finally, after visual inspection, 54 systems were classified into D, 64 into SD and 145 into D/SD. Therefore, the final catalog, presented in Table~\ref{Tab:T1a}, contains 3456 D (85\%), 449 SD (11\%), and 145 EBs (4\%) with uncertain classification (D/SD). 
 
\begin{deluxetable}{CCCCC}

\caption{Confusion matrix of the SVM classifier on 2000 synthetic test EBs} \label{Tab:confMat}

\tablehead{\colhead{} &
\colhead{D} & \colhead{SD} & \colhead{OC} & \colhead{ELL} 
}
\tablewidth{700pt}
\tabletypesize{\scriptsize}
\startdata
D & 525 & 1     & 0   & 1\\
SD & 3  & 411   & 8   & 77\\
OC & 0  & 4     & 470 & 2\\
ELL& 1  & 22    & 4   & 471\\
contam. & 0.01 & 0.06 & 0.02 & 0.14\\
\enddata

\end{deluxetable}

\section{Long-Term Variations} \label{subsec:Long term variations}
\begin{figure*}[htb!]
\plotone{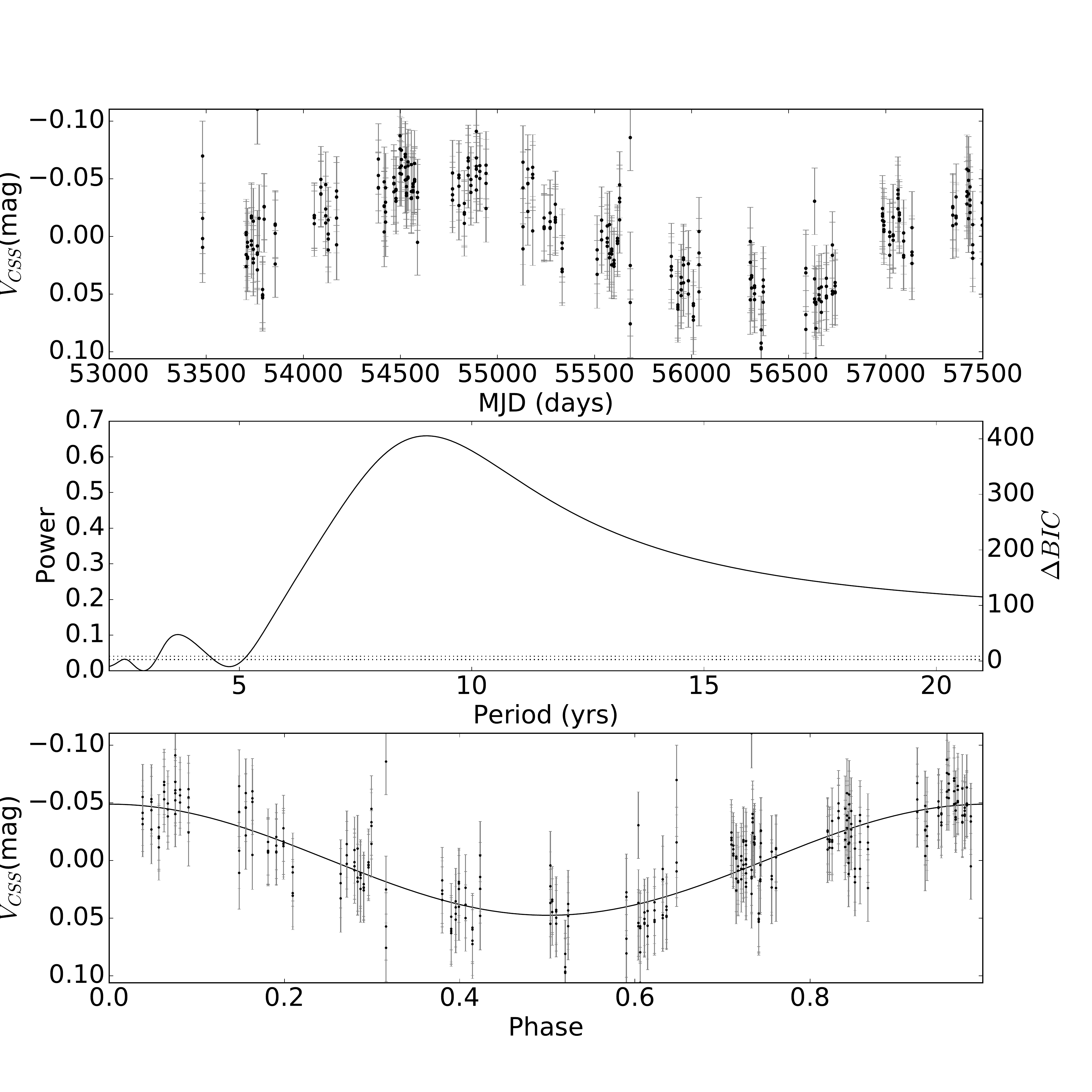}
\caption{Representative example of an EB (CSS J083938.7-050614) showing long-term variation in maximum light. Show here is the result of applying a GLS periodogram to the residuals after subtracting the phenomenological model. {\em Top}: The residuals of the time series data after subtraction of the TGM phenomenological model. {\em Middle}: GLS periodogram analysis of the residuals as a function of time. Dashed lines represent the $1\sigma$ and $3\sigma$ significance levels derived from 1000 Monte Carlo re-samplings. {\em Bottom}: Residuals of the time series data phased with the period derived from the periodogram. \label{fig:f6}}
\end{figure*}

\begin{figure*}[htb!]
\plottwo{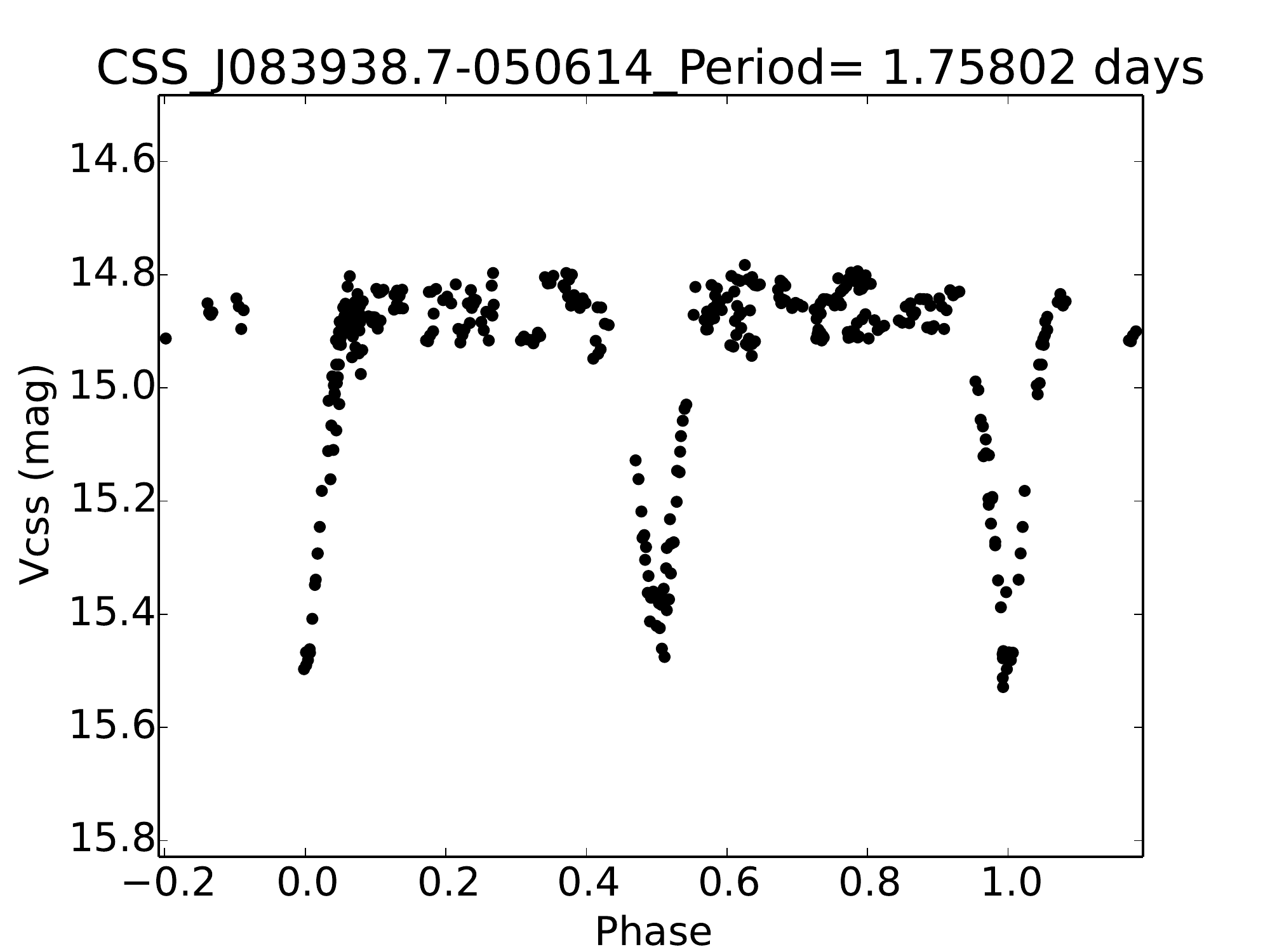}{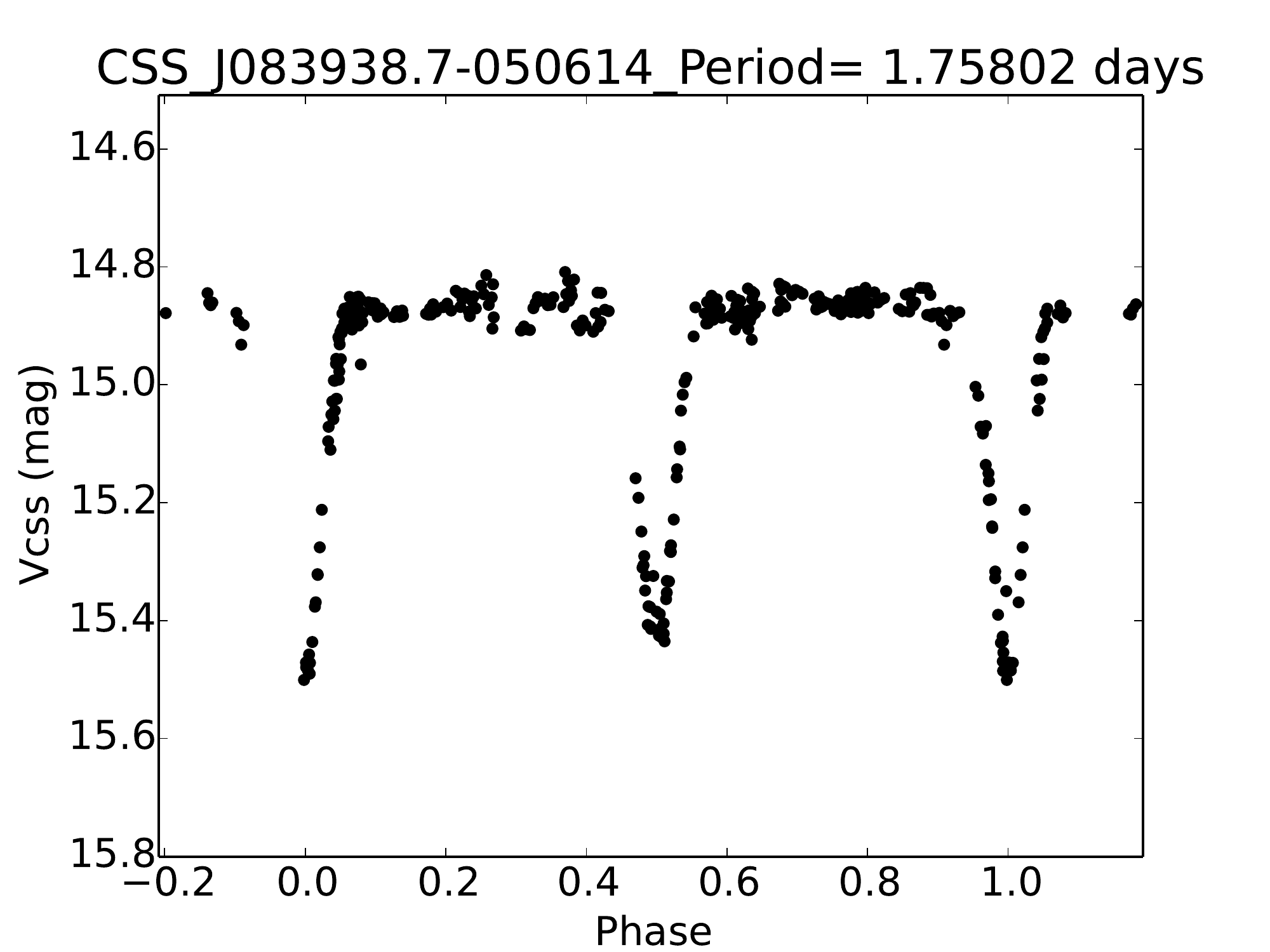}
\plotone{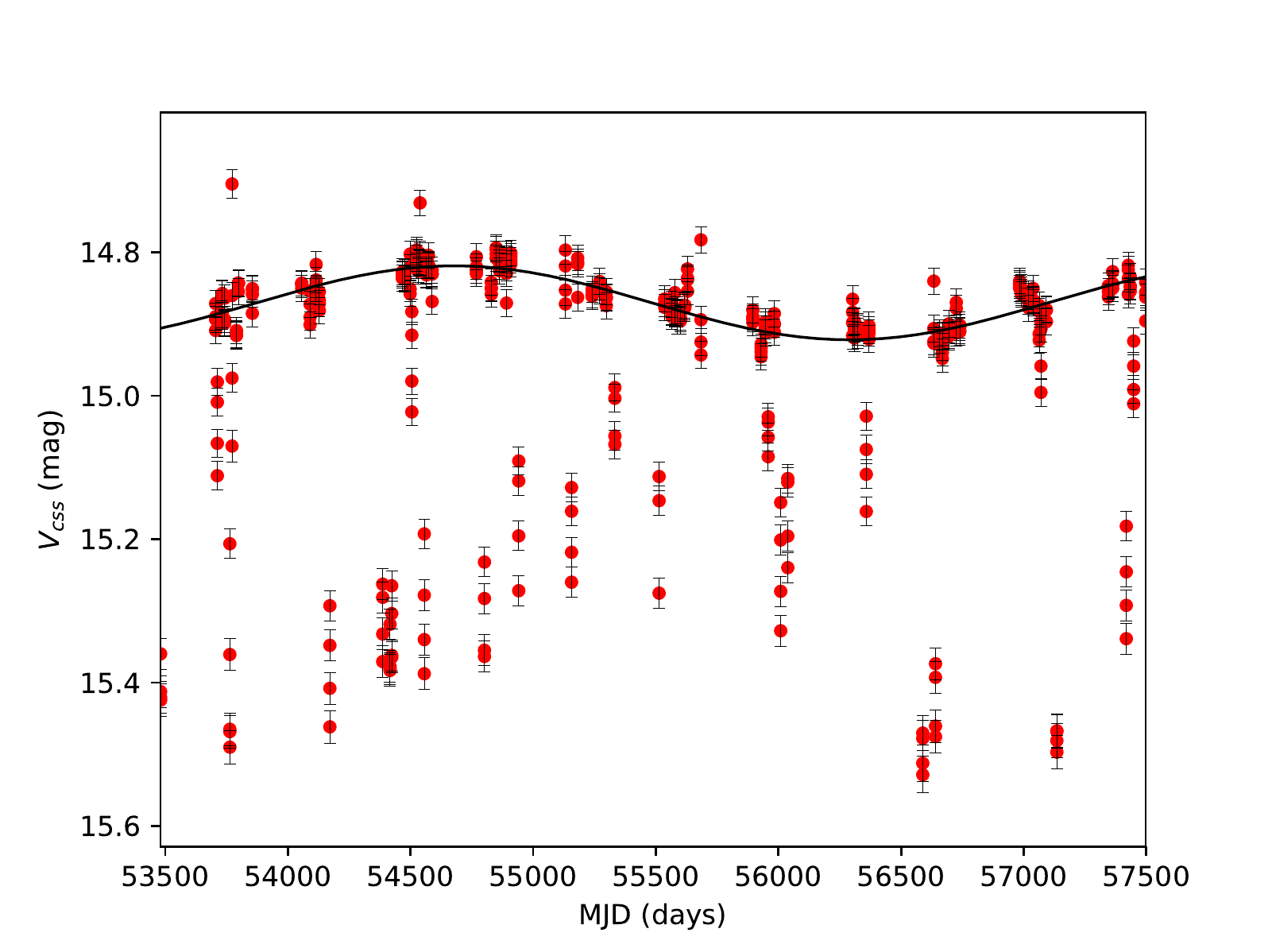}
\caption{{\em Top}: For the same star as in Figure~\ref{fig:f6}, we show here the light curve before ({\em left}) and after ({\em right}) removal of the long-term trend. {\em Bottom}: A harmonic fitting (solid line) was performed on the maximum light after the removal of the eclipses using the method described in Sect.~\ref{subsec:Long term variations}.\label{fig:f5}}
\end{figure*}

Many phased LCs revealed scattering around maximum light, i.e., different maxima in brightness, as shown in Figures~\ref{fig:f6} and \ref{fig:f5}. This made us search for possible long-term changes over the 12~yr timespan of observations.

To detect such variations, we applied three methods. In the first one (Method 1) we subtracted the TGM phenomenological model from the time-series observations (Figure~\ref{fig:f6}, $\it{top}$) and performed a GLS analysis of the residuals (Figure~\ref{fig:f6}, $\it{middle}$), in order to evaluate the possible presence of periodicity in this variation (Figure~\ref{fig:f6}, $\it{bottom}$). For the second and third methods, prior to the fitting, the LCs were binned in time, with bin sizes that depend on the dynamical range of observations, and the median value and standard deviation were calculated for each such bin. Then the eclipses were removed by selecting the data points in the neighborhood of the median values, applying 1$\sigma$ tolerance (Figure~\ref{fig:f5}, {\em top right}). The amplitude and the period of binned LCs were calculated through a GLS periodogram by using the FATS library (Nun et al. 2015) in Python (Method 2) or by applying a harmonic fit to the binned data (Method 3). In order to detect significant variations over long ($\sim 5-10$~yr) timescales, the following set of constrains was applied to the results of the above methods:

\begin{enumerate}
\item LCs with amplitudes of the maxima variation lower than the LC mean error were rejected;
\item LCs with periods of maxima variation $\lesssim 800$~days or $\gtrsim 7000$~days were rejected, due to the available time span of the observations. The upper limit is set by the fact that the total time baseline of the current sample of data of CSS survey is about 12~yr. Thus the period of any parabolic variation must be less than roughly $1.5 \times 12$~yrs;
\item Only signals with Bayesian Information Criterion (BIC, Schwarz 1978) greater than 15 were accepted; 
\item Peak GLS power must be five times the $3\sigma$ power as predicted by 1000 Monte Carlo re-samplings (VanderPlas et al. 2012, 2014; Marsh et al. 2017).
\end{enumerate}

Combining the results from the previous methods and applying the aforementioned criteria, we found 152 systems in the sample of 4680 EBs that appear to exhibit variability in their maxima. For all these systems, the variability seems to be either periodic or quasi-periodic, over long ($\sim 5-10$~yr) timescales. Figure~\ref{fig:f5} ({\em bottom}) shows a representative example with the derived sinusoid model (Method 3) fitted on time-binned data.

The resulting sample was examined for the possibility of the presence of SDSS (Ahn et al. 2012; Alam et al. 2015) sources within $5\arcsec$ of our systems, as any such nearby sources could contaminate the CSS photometry and thus produce spurious variations in the LCs. As a result, 33 EBs were removed from the sample, resulting in 119 EBs with maximum light variation. This final sample of 119 EBs showing long-term variations in the maximum light are labeled in Table~\ref{Tab:T1a} as ``LongTerm +.'' 
 
\subsection{Applegate mechanism vs spot activity} \label{sec:LTmechanism}
Maximum light variations in EBs can be explained by the Applegate mechanism (Applegate 1992) that relates the orbital period modulation to the operation of a hydromagnetic dynamo in the convection zone of the active star in a close binary system. As the active star progresses through its magnetic activity cycle, a changing differential rotation modifies its shape, changing the gravitational quadrupole moment that manifests itself through a cyclically varying orbital period and luminosity of the star, with the same period as the magnetic cycle. Fractional luminosity variations of $\Delta L/L \sim 0.1$ of the active(s) star(s) can produce period variations of $\Delta P/P \sim 10^{-5}$. The majority of our 119 candidates matching Two Micron All-Sky Survey \citep[2MASS;][]{msea06} sources have colors $J-H > 0.237$~mag and $H-K >0.063$~mag, which implies effective temperatures likely lower than $\sim 6200$~K (Pecaut \& Mamajek 2013). Since our results are in agreement with the changes expected under the Applegate mechanism, the latter cannot be excluded as an explanation of the detected maximum light variations. However, if the convective zone cannot respond fast enough (i.e., the thermal timescale of the envelope is much longer than the timescale of the activity cycles), the heat flow variations will be dumped, and thus become unobservable (Watson \& Marsh 2010; Khaliullin \& Khaliullina 2012).
 
On the other hand, maximum light variations could be explained by cool starspot coverage due to the magnetic activity. Our Sun shows such variations in a cycle of 11 yrs, and in case of low-mass EBs where the components are rotating nearly 100 times faster than the Sun, the deep convective envelope along with rapid rotation can produce a strong magnetic dynamo and solar-like magnetic activity. Long lightcurves shed light on the nature of stellar activity of solar or late-type stars, either as single or as members of binary systems (such as BY~Dra and RS~CVn stars; Lehtinen et al. 2016). Marsh et al. (2017) suggested that the variation of the overall brightness in W~UMa-type stars is probably caused by groups of starspots, rather than individual starspots.
 
\begin{figure*}[htb!]
\plottwo{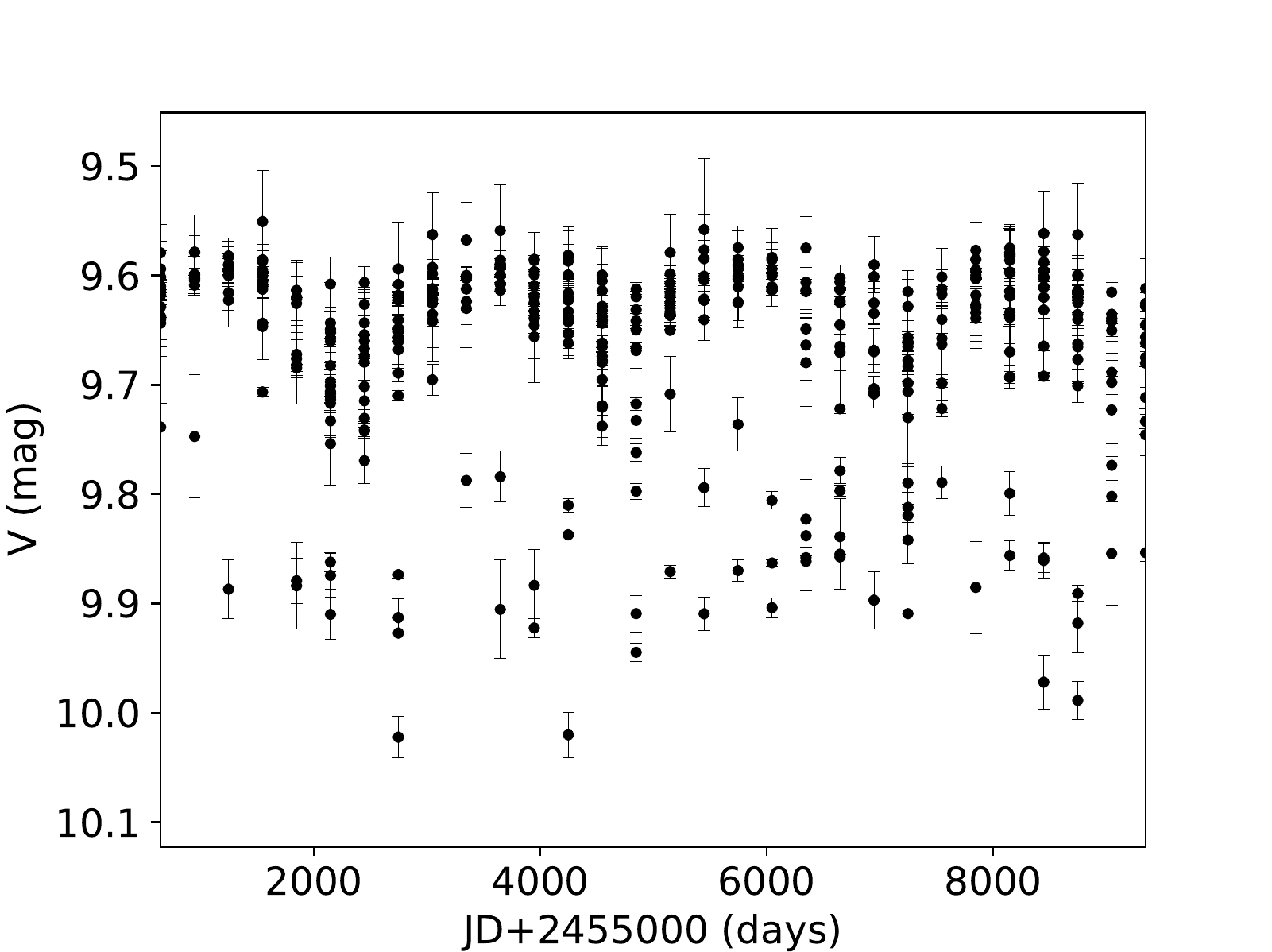}{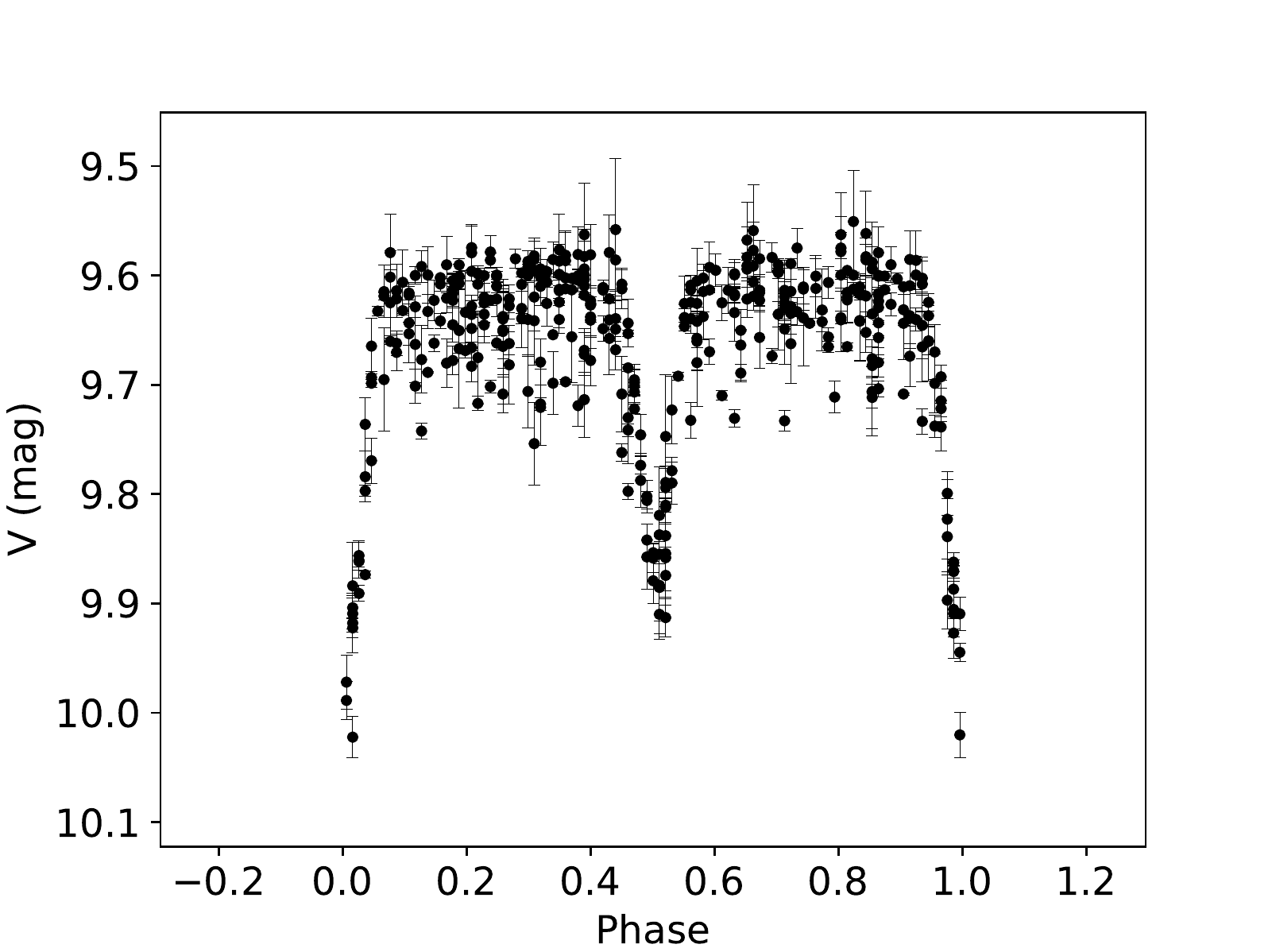}
\caption{Time series ({\em left}) and phase-folded data ({\em right}) of a simulated spotted EB with a magnetic cycle of 6.3~yrs. \label{fig:f11}}
\end{figure*}

In order to simulate the effect of maximum light variations due to starspots, a synthetic eclipsing binary was constructed using the PHOEBE-2.0 engine (Pr{\v s}a et al. 2016) with two main sequence stars with effective temperatures 5880~K and 5490~K. The inclination of the system was set to 80~deg. We assumed four starspot regions in order to simulate a uniform starspot coverage. The positions (longitude and colatitude) of the starspot regions were randomly drawn from a uniform distribution, and the temperature ratio of the spot over the star was set to 0.9 for the primary (active) star. In order to simulate the magnetic cycle, we assumed cyclic variations of the starspot radius (large enough to mimic starspot regions close to maximum activity, and small for lower magnetic activity) in the range of $[0,35]$~deg. A period of 6.3~yrs was assumed for the magnetic cycle. For the virtual observations, a cadence of 300~days and a total timespan of 9000~days were assumed (i.e. we assumed that the virtual observations were carried out in one night every 300~days). Furthermore, variable random noise was added to the time series data, and finally 300 random points were selected for the final simulated light curve. The results of the simulation, under our assumptions (variable size of cool starspot regions), support the explanation of starspot activity (Figure~\ref{fig:f11}). However, in order to achieve amplitudes of variation comparable to our observations and to cover regions that could produce variations also in out-of-eclipse phases, the starspot regions must be large~-- otherwise, we have to assume that both components show magnetic activity. 
  
Both the Applegate mechanism and starspot activity share the same period of the magnetic cycle of the magnetically active star. Accurate times of minimum light observations and period variation analysis are needed to investigate which mechanism(s) may be the underlying cause of these variations. 

\section{Low-Mass Eclipsing Binaries in Catalina Sky Survey} \label{sec:Low mass EBs}
Low-mass EB systems are interesting for determining the fundamental parameters of low-mass stars, which are the most common type of star in the Universe. However, recent studies have shown that they represent significant challenges to the theoretical stellar models, due to their inflated sizes, magnetic activity, and also the poorly understood way they evolve in close binary systems (Chabrier et al. 2007; Feiden 2015; Zhou  et al. 2015). Our list of 4050 classified EBs enable us to search for low-mass EB candidates by imposing color criteria. Accordingly, here we apply the following cuts: $V-K_{s}> 3.0$, as suggested by Hartman et al. (2011); $0.35 <J-H< 0.8$~mag, $H-K_{s} \leq 0.45$~mag, based on L{\'e}pine \& Gaidos (2011) and Zhong et al. (2015).

\begin{figure*}[htb!]
\plotone{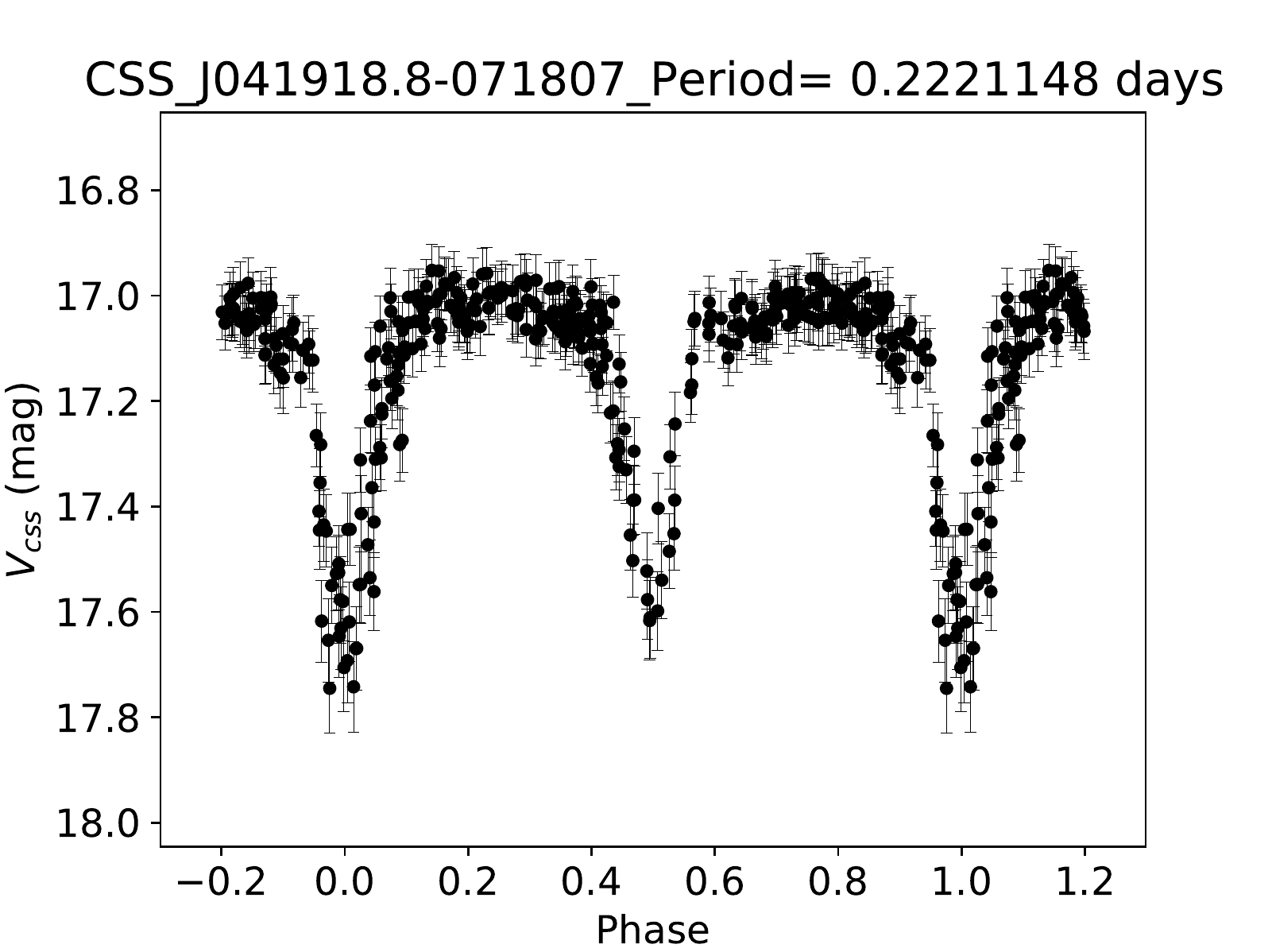}
\caption{Folded LC of CSS J041918.8-071807, the low-mass EB candidate with the shortest period ($P=0.22$~days) in our sample.\label{fig:f7}}
\end{figure*}

\begin{figure*}[htb!]
\plotone{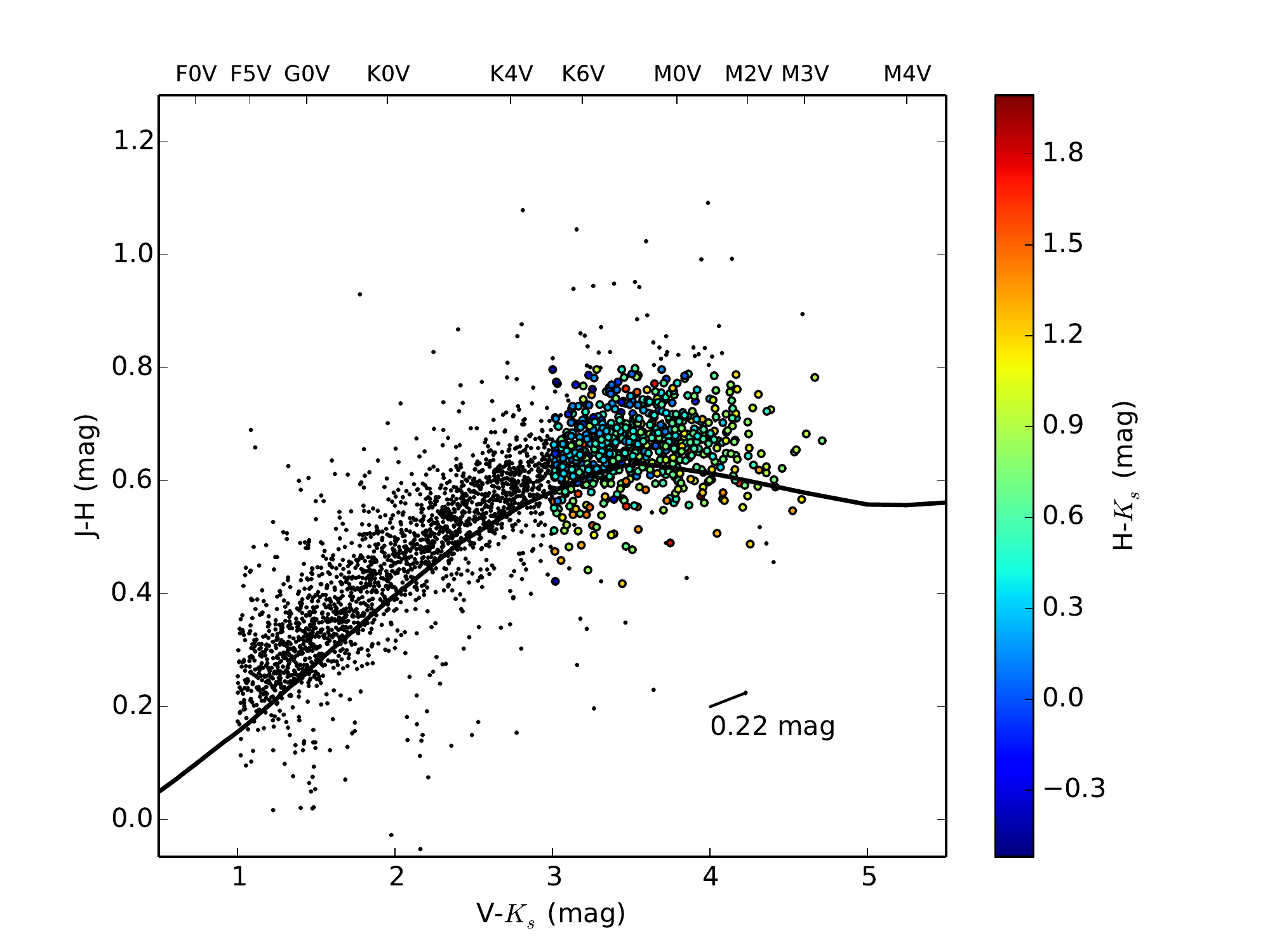}
\plotone{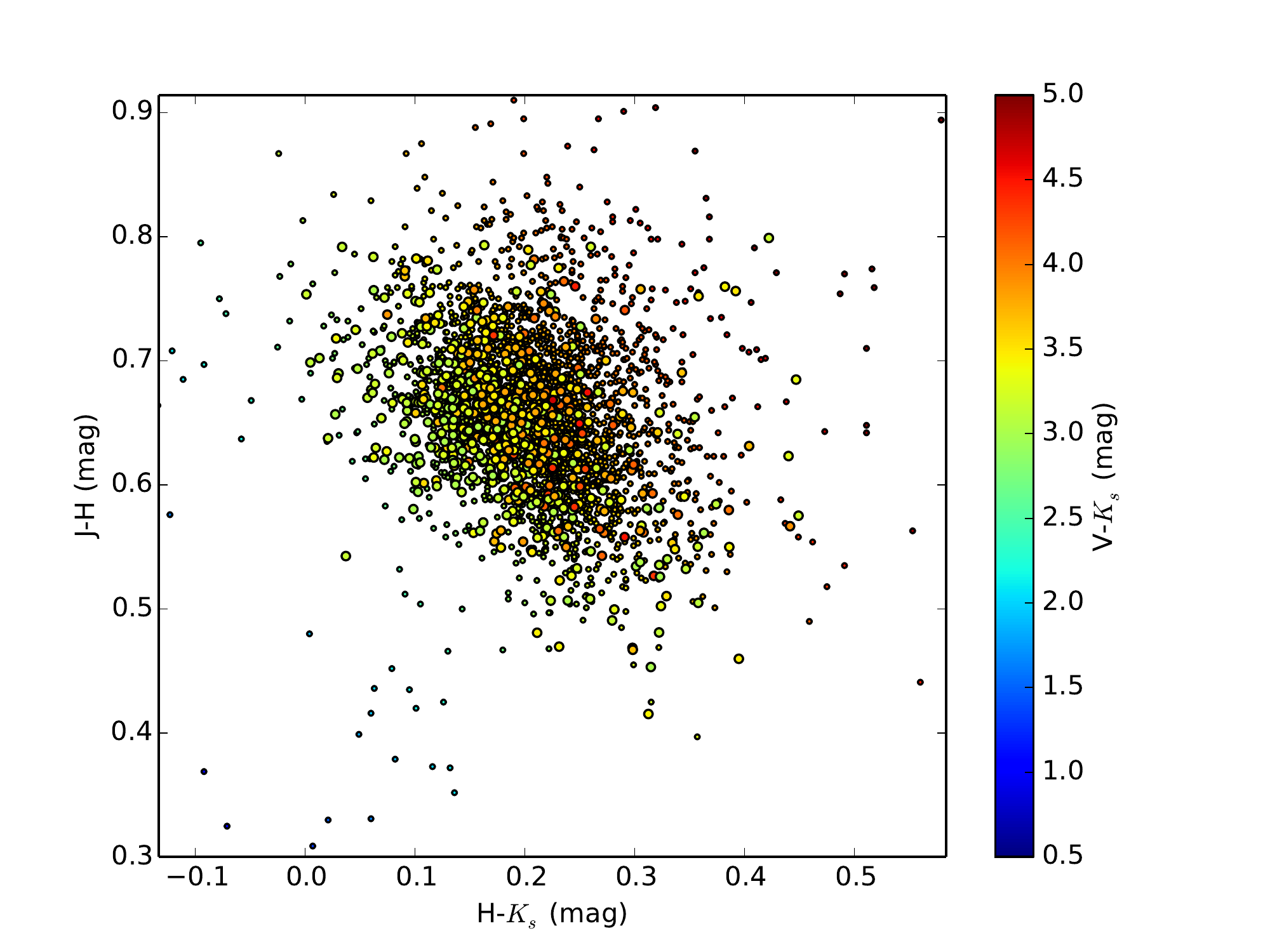}
\caption{{\em Top}: $(V-K_{s}) - (J-H)$ color-color diagram of the 3456 CSS EBs classified as D and the theoretically expected colors of main sequence F5-M3 stars (Pecaut \& Mamajek 2013). Large dots refer to the low-mass EB candidates. The reddening vector was calculated from the mean value of the extinction of the entire sample, while the range is within [0.01-0.59]~mag. {\em Bottom}: $(H-K_{s}) - (J-H)$ color-color diagram of 609 low-mass EB candidates (large dots) overplotted on the sample of low-mass stars from the LAMOST survey (smaller dots). In both panels, the different colors indicate the color index value according to the adjacent color bar. \label{fig:f9}}
\end{figure*}

\begin{figure*}[htb!]
\plotone{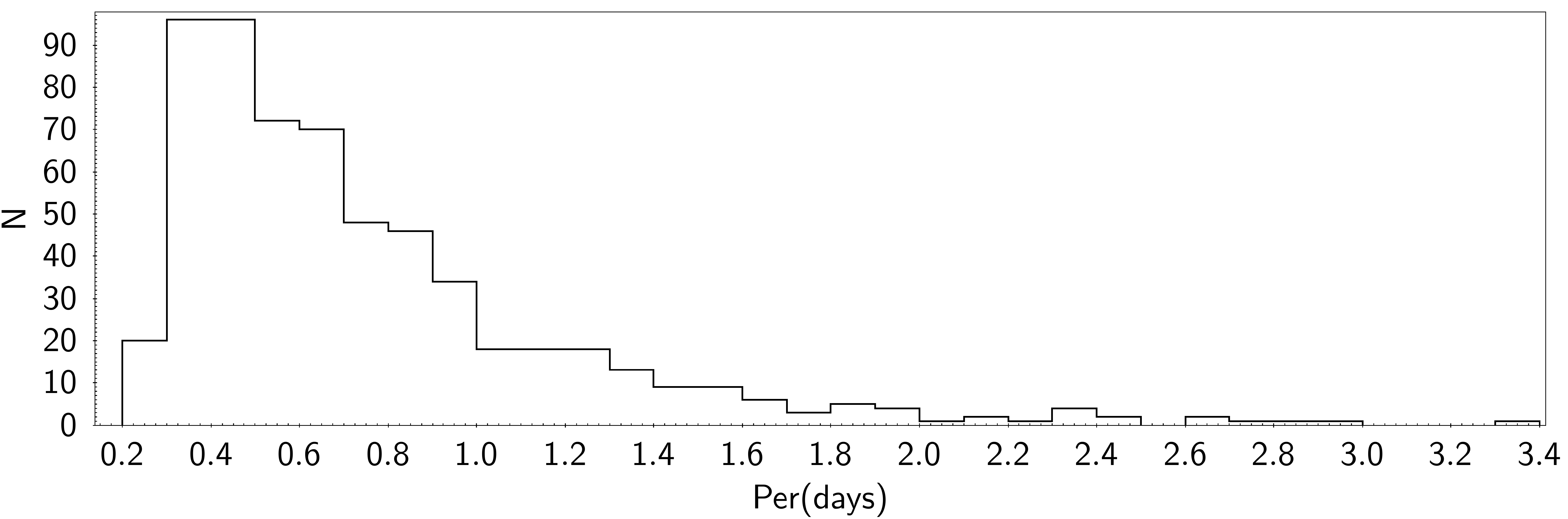}
\caption{Period distribution of 609 low-mass EB candidates identified in this study.\label{fig:f10}}
\end{figure*}

For the color selection, we again use the 2MASS $JHK_{s}$ photometry, performing a cross-match to the 2MASS catalog (Cutri et al. 2003) within 3\arcsec\ of the positions of our stars. Where available, photometry from the APASS survey (Henden et al. 2016) was also used, to obtain the $(B-V)$  color index  in order to transform the $V_{\rm CSS}$ magnitudes to Jonhson $V$. This was accomplished using the same transformation formula as presented in Drake et al. (2013):

\begin{equation}
V = V_{\rm CSS} + 0.31 \times (B - V)^{2} + 0.04.
\end{equation}

This selects 2377 EBs. For the rest of the cataloged EBs for which we have no visual color information, we apply the transformation from 2MASS indices to the Johnson-Cousins system provided by Bilir et al. (2008, their Eq.~16). Interstellar extinction corrections were applied to the $(B-V)$ color index of each EB using the $E(B-V)$ values from Green et al. (2015). The $VJHK_{s}$ magnitudes were also corrected accordingly, using extinction models obtained from a combination of Marshall et al. (2006), Green et al. (2015), and Drimmel et al. (2003), as included in the Python package mwdust\footnote{\url{https://github.com/jobovy/mwdust}} (Bovy et al. 2016). Combining the results with the classification from Section~\ref{subsec:Classification}, only the systems classified as D were finally accepted, resulting in 609 candidates. For distances between 1 and 4.5~kpc, the results are independent of the reddening corrections applied. The periods of the final sample are in the range of $[0.2-3.5]$~days (Figure~\ref{fig:f10}). 

To verify whether these are all bona-fide low-mass EB candidates, we performed two tests:
 
\begin{enumerate}
\item We compare their infrared colors with the theoretically expected colors of main sequence F5-M3 dwarfs, as reported by Pecaut $\&$ Mamajek (2013). The results are shown in the upper panel of Figure~\ref{fig:f9}. The reddening vector in this plot was calculated from the mean value of the extinction of the entire sample; 
\item We compare their infrared colors with the colors of stars in the largest K and M dwarf spectroscopic sample (2612 binaries) from the Large Sky Area Multi-object Fiber Spectroscopic Telescope (LAMOST, Zhong et al. 2015). The results are shown in the lower panel of Figure~\ref{fig:f9}. 
\end{enumerate}

As we can see, the large majority of our sample does indeed fall within the K5 and M3 subtypes. These 609 binaries selected as low-mass candidates are marked in Table~\ref{Tab:T1a} as ``LMcand +''. Examination of Table~1 in Lee (2015) reveals a total of 572 EB systems with nominal components masses $< 0.6 \, M_{\odot}$. However, as explained in that table's header, this includes a large number of systems with uncertain solutions, and also systems with large errors ($> 0.2 \, M_{\odot}$) in the mass values. For this reason, to carry out a meaningful comparison between our results and those reported by Lee (2015), we restrict ourselves to systems with masses in the range covered by our sample (i.e., with spectral types later than K5, or masses $< 0.71 \, M_{\odot}$) and with errors $< 0.1 \, M_{\odot}$. This leads to a total of 107 systems, including 7 EBs classified as either non-detached or uncertain by our analysis, but which Lee classifies as detached. Out of the remaining 100, 72 were matched with our low-mass sample. The remaining 28 systems fall outside the limits of our color selection criteria, thus suggesting that they may not be bona-fide low-mass EB systems. Note that four systems that have been verified as double-lined M-dwarf EBs (Lee \& Lin 2017; Lee 2017); all of them are included in our sample of 609 candidates, but only two appear in Lee's (2015) catalog.

In addition, in our low-mass sample of detached EBs we found candidates near the short-period cutoff at $P \sim 0.22$~d \citep{sr92,sr97}, as can be seen from Figure~\ref{fig:f7}. Only a few such systems are currently known (Drake et al. 2014b). To our knowledge, the detached system with main sequence components with the shortest period known (0.1926~d) is GSC 2314-0530 (=~1SWASP J022050.85+332047.6), identified by Norton et al. (2007) and modeled by Dimitrov $\&$ Kjurkchieva (2010). Nefs et al. (2012) spectroscopically confirmed a detached system with a 0.18~days period containing an M dwarf, but without measuring radial velocities.

\section{Conclusions} \label{sec:conclusion}
Using CSS data covering a 12-yr timespan, we obtained an updated catalog of 4680 EA-type EBs, with revised period determination, phenomenological parameters of their light curves, and system morphology classification based on machine learning techniques. Our study includes many low-mass EB candidates, as well as systems that show additional variation in their maxima over long ($\sim 5-10$~yr) timescales. Most of the new periods are in excellent agreement with those provided in the original Catalina catalogs, but significantly revised values have been obtained for $\sim 10\%$ of the stars. A total of 3456 EBs were classified as D, 449 as SD, and 145 EBs had an uncertain classification. Our classification agrees with the findings of Lee (2015) for 83\% of the sources.
 The sample classified as SD contains $\sim 9\%$ systems with spectral types earlier than F0V, thus it seems that the majority of the systems in the sample are F-G spectral type EA systems with periods less than a day. These systems have been characterized as short period Algols \citep{2000MNRAS.315..587R} in the scenario of \cite{2006AcA....56..199S}.  At the same time, they have also been described as being in near \citep{2001AJ....122.3436V, 2004PASP..116..931K} or broken  \citep{2001AJ....122.3436V,2013AN....334..860A} contact. We again caution, as we did in the Introduction, that a detailed physical modeling of individual EBs is needed to reveal the true system configuration.

Following our methodology of searching for K- and M-type dwarfs, we ended up with a sample of 609 low-mass EB candidates, increasing the total sample of stars at the low-mass end. Spectroscopic follow-up of these sources would be useful to help place constraints on models of low-mass stars. The majority of Lee's (2015) low-mass candidates are included in our sample, including four that have been verified as double-lined M dwarf EBs (Lee \& Lin 2017; Lee 2017). Moreover, we identified rare EA systems with periods close to the period cut-off at $P \sim 0.22$~d \citep{sr92,sr97}.

In addition to these results, our analysis of the long-term trends in the CSS data revealed cyclic or quasi-cyclic modulation of the maximum brightness on long ($\sim 5-10$~yr) timescales for as many as 119 EA systems (2.5\% of the entire sample). The $\Delta L/L$ range is within $[0.04-0.13]$ with mean value $\langle\Delta L/L \rangle = 0.075\pm 0.017$, while the periods are in the range of $[4.5-18]$~yrs, with mean $P =12.1 \pm 3.3$~yrs.

Recently, Marsh et al. (2017) reported similar behavior in 205 eclipsing W~UMa-type systems from CSS (2.2\% of the target sample), finding periods in the range $4-11$~yrs and fractional luminosity variance $\Delta L/L \approx 0.04-0.16$. Close binaries are known to be significantly more active than wide binaries and single stars (e.g., Shkolnik et al. 2010), most likely due to their being tidally locked and high rotational velocities, resulting in high levels of magnetic activity. This in late types is predicted to inflate their radii by inhibiting convective flow and increasing starspot coverage. The observed long-term variability can be explained by  either the Applegate mechanism or by variable spot regions. Ol{\'a}h (2006) suggested that the magnetic field interaction has more effects on the starspot activities of the main-sequence stars than does the tidal force, because these stars have much higher surface gravities. As a consequence, the main-sequence stars often show active regions at quadrature phases. It should be noted that even though the vast majority of spotted stars cannot be easily imaged with special techniques (Doppler imaging or interferometry), our sample is useful to the future study of stellar activity cycles or other associated phenomena (e.g., flares).

\acknowledgements
A.P. and M.C. gratefully acknowledge the support provided by Fondecyt through grants \#3160782 and \#1171273. Additional support for this project is provided by the Ministry for the Economy, Development, and Tourism's Millennium Science Initiative through grant IC\,120009, awarded to the Millennium Institute of Astrophysics (MAS); by Proyecto Basal PFB-06/2007; and by CONICYT's PCI program through grant DPI20140066. M.C. gratefully acknowledges the additional support provided by the Carnegie Observatories through its Distinguished Scientific Visitor program. 
The Monte-Carlo script in PHOEBE-scripter is based on the script that was kindly provided by Dr. Andrej Pr{\v s}a.

This work made use of data products from the CSS survey. The CSS
survey is funded by the National Aeronautics and Space Administration under Grant No. NNG05GF22G issued through the Science
Mission Directorate Near-Earth Objects Observations Program. The
CRTS survey is supported by the US National Science Foundation under grants AST-0909182, AST-1313422, AST-1413600, and
AST-1518308.

This publication makes use of data products from the Two Micron All Sky Survey, which is a joint project of the University of Massachusetts and the Infrared Processing and Analysis Center/California Institute of Technology, funded by the National Aeronautics and Space Administration and the National Science Foundation.
Funding for SDSS-III has been provided by the Alfred P. Sloan Foundation, the Participating Institutions, the National Science Foundation, and the U.S. Department of Energy Office of Science. The SDSS-III web site is http://www.sdss3.org/.

This publication makes use of data products from SDSS-III.
SDSS-III is managed by the Astrophysical Research Consortium for the Participating Institutions of the SDSS-III Collaboration including the University of Arizona, the Brazilian Participation Group, Brookhaven National Laboratory, Carnegie Mellon University, University of Florida, the French Participation Group, the German Participation Group, Harvard University, the Instituto de Astrofisica de Canarias, the Michigan State/Notre Dame/JINA Participation Group, Johns Hopkins University, Lawrence Berkeley National Laboratory, Max Planck Institute for Astrophysics, Max Planck Institute for Extraterrestrial Physics, New Mexico State University, New York University, Ohio State University, Pennsylvania State University, University of Portsmouth, Princeton University, the Spanish Participation Group, University of Tokyo, University of Utah, Vanderbilt University, University of Virginia, University of Washington, and Yale University. 
This work has made use of the SIMBAD database, operated at CDS, Strasbourg, France.
 
This research was made possible through the use of the AAVSO Photometric All-Sky Survey (APASS), funded by the Robert Martin Ayers Sciences Fund.
\vspace{5mm} 
\software{AstroML \citep{2012cidu.conf...47V},
          CKP \citep{2015ApJS..216...25P},
          FATS \citep{Nun15},
          LMFIT \citep{2016ascl.soft06014N},
          mwdust \citep{	2016ApJ...818..130B},
          PDM \citep{1978ApJ...224..953S},
          PHOEBE-scripter \citep{Pr05},
          PHOEBE-2.0 \citep{2016ApJS..227...29P},
          pyMC \citep{2015ascl.soft06005F},
          scikit-learn \citep{2012arXiv1201.0490P},
          triangle.py-v0.1.1 \citep{2014zndo.soft11020F},
          VARTOOLS \citep{2016A&C....17....1H} 
          }
\appendix

\section{ACRONYMS AND ABBREVIATIONS}

\begin{deluxetable}{cc}[htb!]
\tablewidth{700pt}
\tabletypesize{\scriptsize}
\tablehead{
\colhead{ } &\colhead{Description} 
}
\startdata
CSS &Catalina Sky Survey \\
VISTA & Visible and Infrared Survey Telescope for Astronomy \\
ASAS &All Sky Automated Survey  \\
NSVS &Northern Sky Variability Survey \\
VVV &Variables in the Via Lactea \\
TrES &Transatlantic Exoplanet Survey TrES \\
OGLE &Optical Gravitational Lensing Experiment \\
HATNet &Hungarian-made Automated Telescope Network exoplanet survey  \\
SuperWASP &Wide Angle Search for Planets  \\
2MASS &Two Micron All-Sky Survey  \\
APASS &AAVSO Photometric All-Sky Survey \\
LAMOST &Large Sky Area Multi-object Fiber Spectroscopic Telescope  \\
CSDR2 &Catalina Surveys Data Release 2 \\
AoV  & Analysis of Variance    \\
BLS & Box-Least Squares   \\
GLS & Generalized Lomb-Scargle  \\
PDM & Phase Dispersion Minimization \\
CKP & Correntropy Kernelized Periodogram \\
MCMC & Markov Chain Monte Carlo  \\
SVM & Support Vector Machine \\
ANN & Artificial Neural Network \\
KN & K-nearest neighbor \\
BIC & Bayesian Information Criterion \\
LM &  Levenberg-Marquardt nonlinear minimization algorithm \\
\enddata
\end{deluxetable}

\end{document}